\documentclass[%
 reprint,
showpacs,
 amsmath,amssymb,
 aps,
]{revtex4-1}
\usepackage{graphicx}
\usepackage{dcolumn}
\usepackage{bm}
\usepackage{color}
\usepackage{enumitem}
\usepackage{textcomp}
\usepackage[bottom]{footmisc} 

\usepackage[
margin=0.65in,top=0.1in,
]{geometry}

\begin{document}

\preprint{APS/123-QED}

\title{Exciton-driven giant nonlinear overtone signals from buckled hexagonal monolayer GaAs}
\author{Himani Mishra}
 \author{Sitangshu Bhattacharya}
 \email{Corresponding Author's Email: sitangshu@iiita.ac.in}
\affiliation{Nanoscale Electro-Thermal Laboratory, Department of Electronics and Communication Engineering, Indian Institute of Information Technology-Allahabad, Uttar Pradesh 211015, India}
\begin{abstract}
We report here a giant $\left|\chi_{baa}^{(2)}\right|=$780 pm/V second harmonic and $\left|\chi_{aaaa}^{(3)}\right|=$1.4$\times$10$^{-17}$ m$^{2}$/V$^{2}$ third harmonic signal from single atomic sheet of buckled hexagonal GaAs. We demonstrate this through the solution of an ab-initio real-time Bethe-Salpeter equation by including the electron-hole screened-exchange self-energy. The coupling between time-dependent external electric field and correlated electrons is treated within the modern theory of polarization. The result of our calculation envisage monolayer GaAs to be a prominent member in the material library for nonlinear signal generations.   
\end{abstract}
\maketitle

\section{Introduction} \label{intro}
Breaking and pairing of symmetry rules in crystalline structure leads to many exciting physical phenomena. For example, the pairing of a broken-inversion and time-reversal symmetry in presence of a laser constructs a direct valley entangled electron-hole recombination that sparks a strong linear (i.e., absorption) as well as a nonlinear optical (NLO) response (i.e., second, third, and other higher harmonic generations). In recent years, there had been benchmark spectroscopic experiments performed on exfoliated and van der Waals epitaxy monolayer (ML) of transitional metal dichalcogenides (TMDCs) \cite{Nardeep2013, Wang2014, Wang2015, Yilei2013, Leandro2015, Ribeiro2015, Henrique2018, Anton2018} and monochalcogenides (TMMCs) \cite{Xu2015, Jiadong2018} demonstrating extraordinary NLO responses. These materials possesses a crystalline non-centrosymmetricity and therefore, the non-cancellation of induced dipole leads to the second harmonic generation (SHG) as a lowest detected NL response. Such responses find wide applications in two-dimensional (2D) optical modulators \cite{Sun2016}, surface morphology characterizations and sum and difference frequency generations \cite{Shen1989, Daria2019}, etc.\\
The frequency dependent induced macroscopic polarization in an NL medium is governed by the infinite series \cite{Franken1961, Boyd2008} $\varepsilon_{0}^{-1}\mathcal{P}_{i}\left(\omega\right)=\left[\chi_{ij}^{(1)}\left(\omega\right)\mathcal{E}_{j}\left(\omega\right)+\right.$ $\chi_{ijk}^{(2)}\left(-\omega;\omega_{1},\omega_{2}\right)\mathcal{E}_{j}\left(\omega_{1}\right)\mathcal{E}_{k}\left(\omega_{2}\right)+$ $\chi_{ijkl}^{(3)}$ $\left.\left(-\omega;\omega_{1},\omega_{2},\omega_{3}\right)\mathcal{E}_{j}\left(\omega_{1}\right)\mathcal{E}_{k}\left(\omega_{2}\right)\mathcal{E}_{l}\left(\omega_{3}\right)+...\right]$ where $\chi_{ij}^{(1)}$ is the linear response, $\chi_{ijk}^{(2)}$ is the SHG response, $\chi_{ijkl}^{(3)}$ is the third harmonic generation (THG) response and so on, and $\mathcal{E}(t)$ is the time-dependent (TD) external electric field. SHG is the condition when $\omega_{1}$=$\omega_{2}$=$\omega$, thus $\chi_{ijk}^{(2)}\left(-2\omega;\omega,\omega\right)$. THG is the condition when $\omega_{1}$=$\omega_{2}$=$\omega_{3}$=$\omega$, thus $\chi_{ijkl}^{(3)}\left(-3\omega;\omega,\omega,\omega\right)$. The index $i$ is the polarization direction while $j$, $k$, $l$, ... are the electric field directions. \\
Within the linear response theory, $\Im\chi_{ij}^{(1)}\left(\omega\right)$ contains rich source of information about the optical absorption spectrum. The peaks in the spectrum with energy below the gap correspond to bound state electron-hole pairs, known as excitons. Both the excitonic ground and excited state energies about the fundamental gap can be obtained once $\Im\chi_{ij}^{(1)}\left(\omega\right)$ is known. Further information on exciton binding energy, shape of the absorption spectrum above the gap and its deviations from the independent particle approximation (IPA) can also be extracted from the photo-luminescence spectroscopy, and therefore such features can be tailored via tuning the excitonic energies and lifetimes.\\
The SHG and THG responses in principle can be obtained experimentally by altering the average incident power (or intensities) of the applied field from the pump and measuring the corresponding reflected average power (or intensities). Once the measurement is done, a suitable physics based model is then used to extract these responses. In the case of 2D materials, two common models: ``sheet" and ``bulk" are used to obtain these responses. The main difference in the SHG responses extracted from these two models is that the refractive index ($n$) of the substrate at $\omega$ appears in the former model, whereas an intrinsic refractive index of the 2D material itself both at $\omega$ and 2$\omega$ appears in the later. Particularly, Clark $et$ $al.$ \cite{Clark2014} demonstrated that the bulk and the sheet SHG responses are related by a ``scaling factor" $F$ (in order of 10) that takes care of the reflectance and transmittance emerging due to the refractive index mismatch between the bulk crystal and the underlying substrate and air, respectively. These authors demonstrated that the pre-factor in $\chi_{bulk}^{\left(2\right)}=32\pi F\frac{n_{2D}\left(\omega\right)\sqrt{n_{2D}\left(2\omega\right)}}{\left(n_{substrate}+1\right)^{3}}\chi_{sheet}^{\left(2\right)}$ is close to 10$^3$ which scales $\left|\chi_{bulk}^{\left(2\right)}\right|$ $\sim$10$^5$ pm/V. This essentially provides a 3-4 orders of magnitude difference between the two models. For instance, the 2D MoS$_{2}$ \cite{Clark2014} and GaSe \cite{Xu2015} reportedly exhibits a giant SHG of the order 10$^{5}$ and 10$^{3}$ pm/V respectively when calculated using the bulk model. In contrast, the sheet model predicts only upto $\sim$100-500 and $\sim$400 pm/V respectively \cite{Clark2014}. The THG response is likewise indirectly evaluated by measuring the reflected average power of the TH beam. A similar sheet model, depending on the refractive index of the substrate at $\omega$, is finally used to obtain the TH response \cite{Anton2018}. The reported THG response for 2D MoS$_2$ and WSe$_2$ is found to be in the order of 10$^{-19}$ m$^{2}$V$^{-2}$ \cite{Wang2014, Henrique2018, Anton2018}. Interestingly, the best THG response in ML family is still from graphene $\sim$ 10$^{-16}$ m$^{2}$V$^{-2}$ \cite{Antti2013, Sung2013}. However, graphene is centrosymmetric and thus the SHG is identically zero.\\
\begin{figure}[ht]
  \includegraphics[width=0.9\columnwidth]{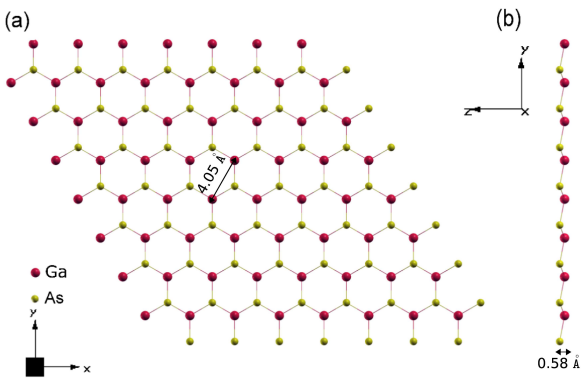}
  \caption{(a) Crystalline structure of monolayer GaAs exhibiting a hexagonal C$_{3v}$ 3m point group symmetry. An in-plane lattice constant $a$=4.05 $\mathring{\mathtt{A}}$ and (b) a buckling height along the out-of-plane (z-direction) is found to be 0.58 $\mathring{\mathtt{A}}$ after performing a ground state energy relaxation calculation.}
  \label{fig:fgr1}
\end{figure}
First principles calculations based on the solution of the many-body GW and Bethe-Salpeter equation (BSE) \cite {Onida2002, Paleari2019} are nowadays the most reliable theoretical approach to investigate exciton affairs in crystals. The excitonic spectra within the linear optics can be achieved from the solution of a time-independent equation of motion (EOM), whereas it needs a solution from the time-dependent EOM to efficiently understand the NL behaviour \cite{Claudio2013}. In-fact, the reduced SHG values mentioned in the preceding paragraph are also supported by the ab-initio calculations \cite{Claudio2019}, thus validating the sheet model to be a more accurate description for SHG and THG in MLs. Using the first principles based calculations and a real-time (RT) approach \cite{Claudio2011} to solve the TD-BSE, we estimate here the SHG and THG responses in ML GaAs. We demonstrate that buckled ML GaAs offers a giant exciton binding energy of 1.10 eV, an SHG of 780 pm/V and a THG of 1.4$\times$10$^{-17}$ m$^{2}$V$^{-2}$. These appealing all-in-one values may open window to let enter ML GaAs as a prominent member in the opto-electronics family. What follows, our approach is based on a fully state-of-the-art ab-initio based calculations and is free from any $ad$-$hoc$ parameters. This paper is organized as follows: In Sec. \ref{compu}, we discuss in separate subsections the detailed methodology of both ground state and excited state linear and nonlinear calculations. This is followed by the results and discussion outlined in Sec. \ref{result} for both the above cases. Finally, Sec. \ref{conclu} briefs the summary of our work.  The Sec. Appendix outlines the necessary mathematical equations which we solved using the ab-initio methods. Additionally, the Supplemental Material \cite{Supplemental} contains the supportive figures and convergence criteria used in this work.
\section{Computational Details} \label{compu}
\subsection{Ground State Calculations}
All density functional theory (DFT) calculations were carried out with Quantum Espresso package \cite{Giannozzi2017}. A fully relativistic, norm-conserving pseudopotential with PBE exchange-correlational functional was used. The 3d semi-core orbital was included in both Ga and As along with 4s and 4p valence electrons. A kinetic cut-off energy of 120 Ry (see Fig. S2 in the Supplemental Material \cite{Supplemental} for details of convergence criteria) was selected. The Brillouin zone (BZ) was sampled on a 12$\times$12$\times$1 grid using a $\Gamma$ centred Monkhorst-Pack scheme with the force and energy thresholds of 10$^{-5}$ Ry/Bohr and 10$^{-5}$ Ry respectively. A vacuum of 30 $\mathring{\mathtt{A}}$ in either side of the ML was selected to prevent the Coulomb interference between the repeated images. A two-spinor wave-function along with the non-collinear and spin-orbit coupling (SOC) criteria was expanded in the plane-wave basis set. The resulted in-plane and buckled lattice constants achieved were $a$=4.05 $\mathring{\mathtt{A}}$ and 0.58 $\mathring{\mathtt{A}}$ respectively (see Fig. \ref{fig:fgr1}). The lattice vibration calculation was carried out on a uniform 18$\times$18$\times$1 dense phonon grid using a rigid self-consistent error threshold below 10$^{-18}$ Ry (see Fig. S6 in the Supplemental Material \cite{Supplemental} for details of phonon convergence criteria).
\subsection{Excited state G$_{0}$W$_{0}$+BSE and real-time BSE calculations}
All linear and nonlinear optical excitations were computed using the extended version of many-body perturbation theory YAMBO package \cite{Sangalli2019, Claudio2013}. 200 bands (28 occupied and 172 unoccupied) along with a response block size of 7 Ry (see Fig. S7 in the Supplemental Material \cite{Supplemental} for details of convergence criteria) were considered in the evaluation of polarization function within the random-phase approximation in the presence of local-field effects. A Godby-Needs \cite{Godby1989} plasmon-pole approximate model was used to calculate the microscopic inverse dynamic dielectric screening function. The transferred momentum \textbf{q} divergences (see Eqn. (\ref{Monte}) in the Appendix) were fixed using a random integration method \cite{Olivia1998, Rozzi2006} in which by keeping the potential unchanged, a smooth momenta integrand function is assumed in each small volumetric region of the BZ. This BZ integral is finally evaluated through a Montecarlo method. 10$^6$ random-\textbf{q} points with a cut-off of 3 Ry were found sufficient to cover and compute the BZ integral fully. Likewise in DFT, a Coulomb truncation of 30 $\mathring{\mathtt{A}}$ on either ML side in form of a box structure was used. The BSE computation was done by dense sampling the BZ to 72$\times$72$\times$1 on a shifted grid and then mapping it to 12$\times$12$\times$1 regular grid. This methodology \cite{Kammerlander2012} essentially speed up the numerical computation without losing the numerical accuracy. The screening part W$_0$ in BSE was obtained by borrowing the statically screened microscopic dielectric function evaluated during G$_0$W$_0$. The full diagonalization of the BS matrix was finally implemented after considering the anti-resonant elements as well. The linear spectrum is found to get converged with top five valence and lowest five conduction bands (see Fig. S8 in the Supplemental Material \cite{Supplemental} for details of convergence criteria). To know the exciton wave-function radius, a hole was kept on the top of As atom at a distance of 1 $\mathring{\mathtt{A}}$. This is indicated by a dark spot in the inset of Fig. \ref{fig:fgr5}. In real-time calculations, a length-gauge was used \cite{Claudio2013} and the subsequent calculations were then carried out with the same 72$\times$72$\times$1 grid and same number of transition bands.
\section{Results and Discussions} \label{result}
\subsection{Ground State and Linear Response}
\begin{figure}[ht]
  \includegraphics[width=1\columnwidth]{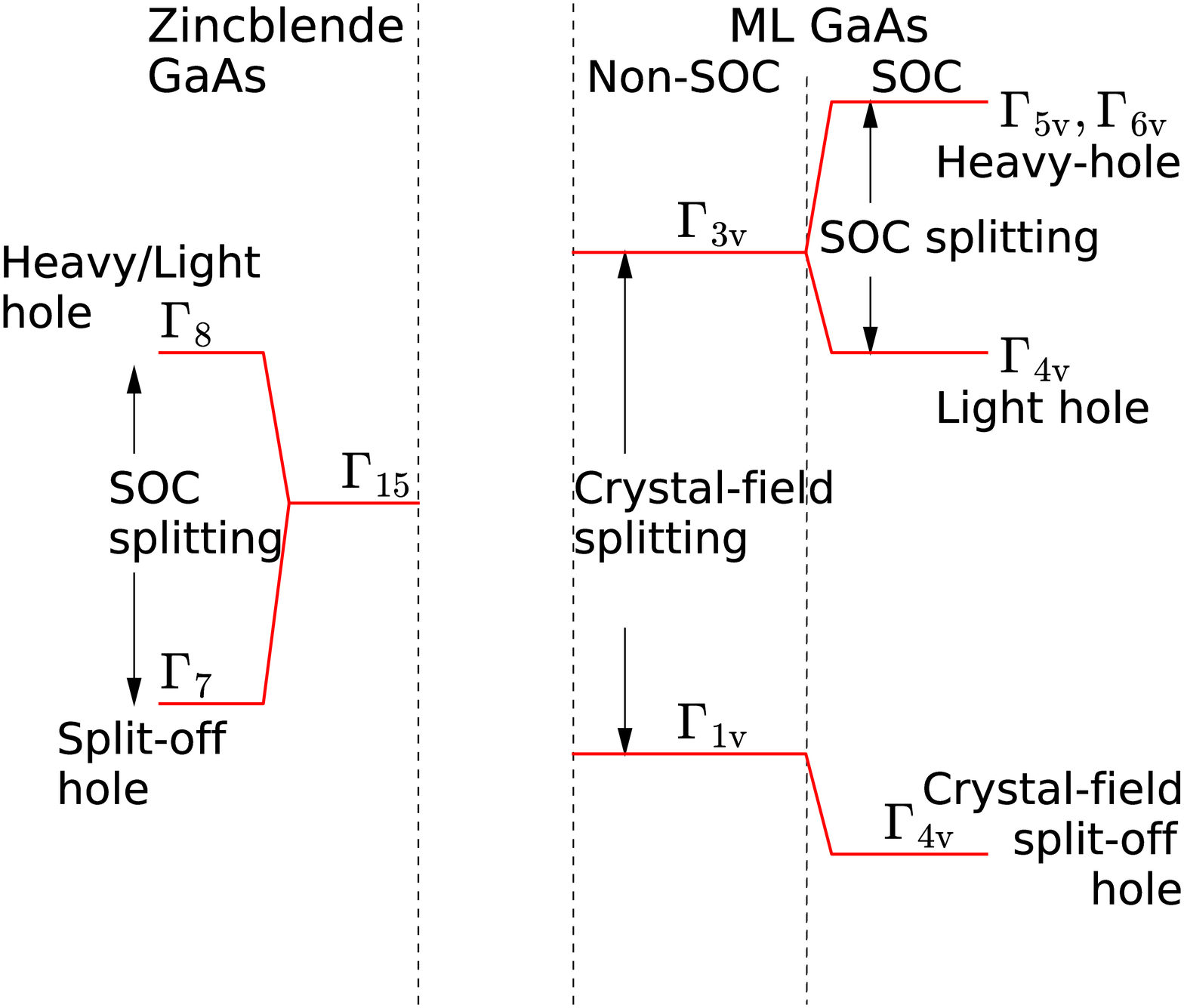}
  \caption{Schematic of valence band splitting at $\Gamma$ of the BZ in the presence of crystal-field and SOC for zincblende and hexagonal GaAs monolayer type structures. Note that $\Gamma_{5\mathrm{v}}$, $\Gamma_{6\mathrm{v}}$ is the Kramers pair with $\left|\mathrm{j},\mathrm{m}_{\mathrm{j}}\right\rangle =\left|\frac{3}{2},\pm\frac{3}{2}\right\rangle $.}
  \label{fig:fgr2}
\end{figure}

\begin{figure*}[!ht]
\includegraphics[width=1.7\columnwidth]{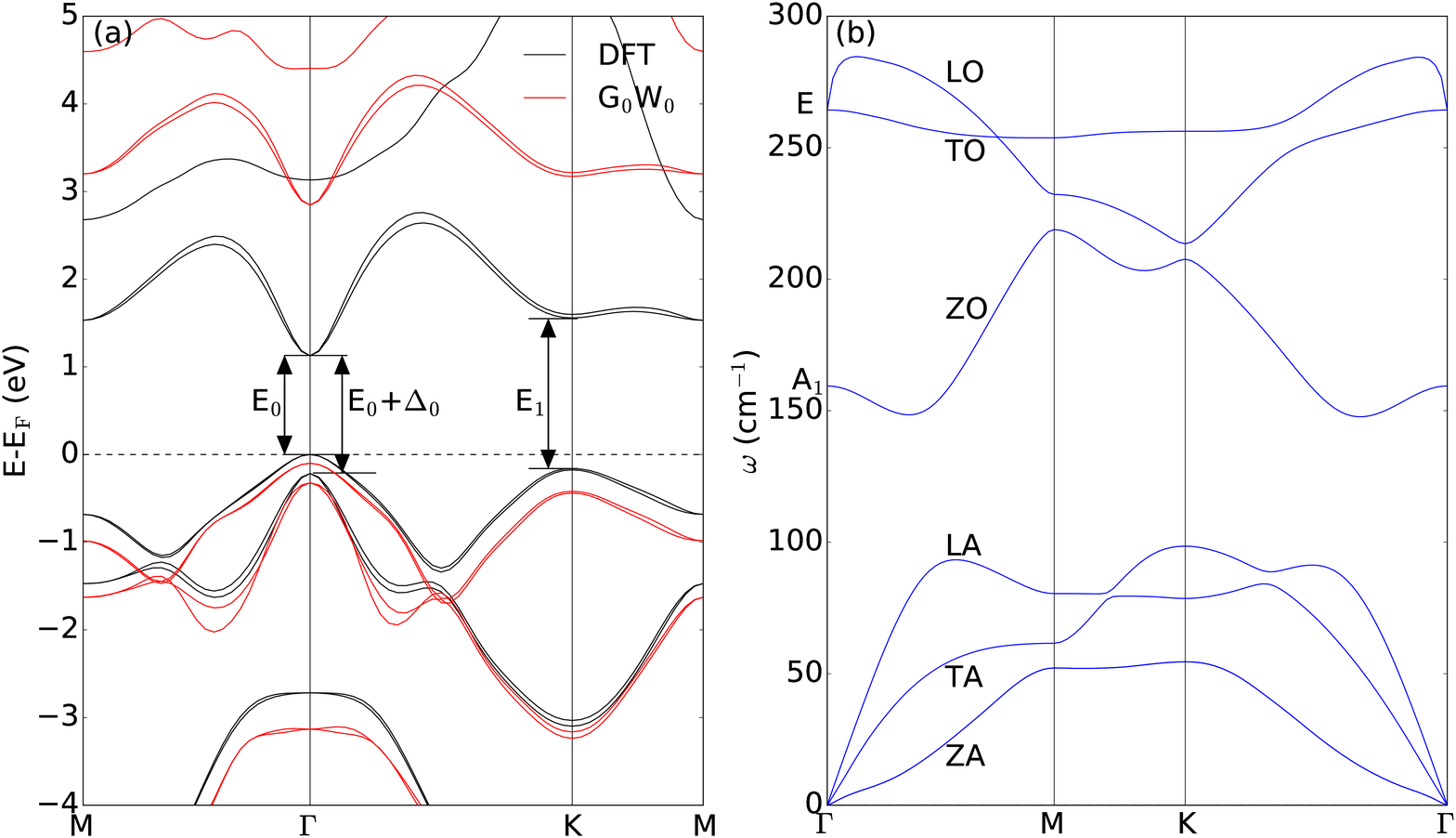}
\caption{(a) Electronic dispersion of ML buckled GaAs using the ground state DFT and excited state G$_{0}$W$_{0}$ theory. The zero-level is set at the top of the DFT valence band. The single-shot GW or G$_{0}$W$_{0}$ improves the gap E$_0$ at $\Gamma$ by 1.82 eV. The valence spin-splitting $\Delta_{0}$ is 0.22 eV. E$_1$ is the gap at \textbf{K}. (b) Phonon dispersion showing no soft modes at $\Gamma$.}
  \label{fig:fgr4}
\end{figure*}

To support our results, we start with the ground state calculations. In contrast to the conventional zincblende and wurtzite crystalline structures, the monolayer GaAs exhibit an in-plane hexagonal lattice symmetry (like a monolayer h-BN) with alternating Ga and As atom and a small buckling height with large interlayer separation. In the bulk zincblende structure, the valence band at the BZ centre exhibits a six-fold degenerate $\Gamma_{15}$ state \cite{Lautenschlager1987}. As schematically shown in Fig. \ref{fig:fgr2}, the presence of an SOC splits this state to a four-fold degenerate $\Gamma_{8}$ heavy and light hole state (higher in energy) and a doubly-degenerate split-off hole $\Gamma_{7}$ (lower in energy) state. In case of single layer hexagonal GaAs, the point group becomes C$_{3v}$ 3m, resulting in a three-fold symmetry (see Fig. \ref{fig:fgr2}). In the presence of crystal-field only (non-SOC), the irreducible representations $\Gamma_{1\mathrm{v}}$, $\Gamma_{2\mathrm{\mathrm{v}}}$ and $\Gamma_{3\mathrm{v}}$ splits into two-fold degenerate $\Gamma_{2\mathrm{v}}$ or $\Gamma_{3\mathrm{v}}$ (higher in energy) and a non-degenerate level $\Gamma_{1\mathrm{v}}$ (lower in energy). In the presence of an SOC, the state $\Gamma_{3\mathrm{v}}$ splits into three levels: $\Gamma_{3v}\otimes\Gamma_{4v}$=$\Gamma_{4\mathrm{v}}$ (doubly-degenerate)+$\Gamma_{5\mathrm{v}}$+$\Gamma_{6\mathrm{v}}$ in which due to the presence of time-reversal symmetry, $\Gamma_{5\mathrm{v}}$ and $\Gamma_{6\mathrm{v}}$ becomes Kramers pair. The crystal-field split lower level $\Gamma_{1\mathrm{v}}$ in presence of SOC becomes $\Gamma_{1\mathrm{v}}\otimes\Gamma_{4\mathrm{v}}$=$\Gamma_{4\mathrm{v}}$ (doubly-degenerate). Consequently, the doublet $\Gamma_{5\mathrm{v}}$, $\Gamma_{6\mathrm{v}}$ corresponds to a heavy-hole, $\Gamma_{4\mathrm{v}}$ to a light-hole and the lower $\Gamma_{4\mathrm{v}}$ (hereafter symbolically represented as $\Gamma_{4\mathrm{v}}^{\mathrm{cf}}$) to a crystal-field split-off hole respectively.
\begin{figure*}[!ht]
\includegraphics[width=2.05\columnwidth]{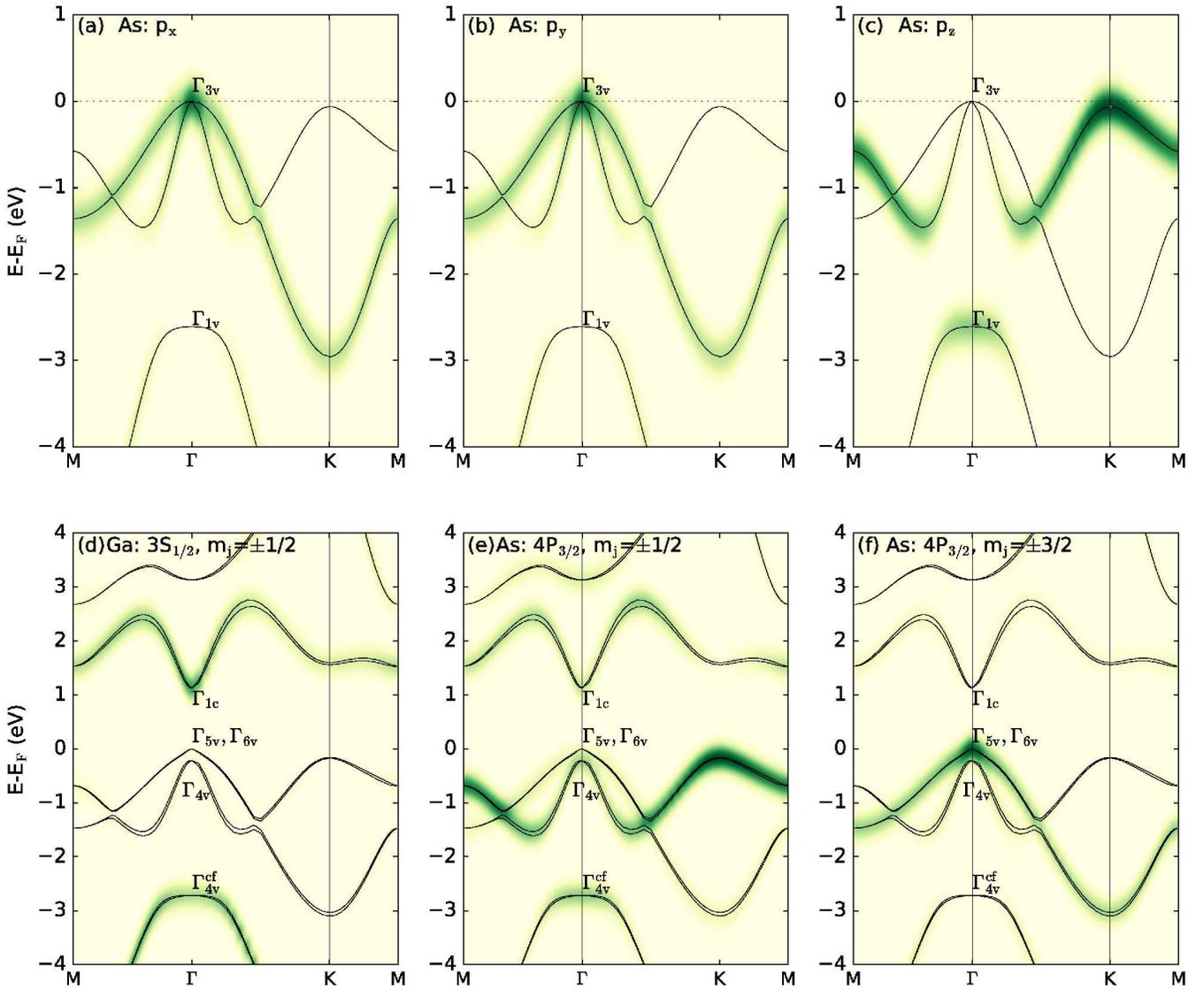}
\caption{Ground state electron energy dispersion in ML buckled GaAs in the absence of SOC demonstrating the p-DOS contributions from (a) p$_{x}$, (b) p$_{y}$ and (c) p$_{z}$ orbitals of As. The difference $\Gamma_{3\mathrm{v}}$-$\Gamma_{1\mathrm{v}}$= -1.95 eV, signifying a strong crystal-field splitting. In the presence of an SOC, the p-DOS contributions for (d) Ga 3S$_{1/2}$, m$_\mathrm{j}$=$\pm$1/2, (e) As 4P$_{3/2}$, m$_\mathrm{j}$=$\pm$1/2 and (f) As 4P$_{3/2}$, m$_\mathrm{j}$=$\pm$3/2 are demonstrated. The presence of SOC makes $\Gamma_{1\mathrm{v}}$-$\Gamma_{4\mathrm{v}}^{\mathrm{cf}}$= -0.11 eV. At the $\Gamma$ point, the valence spin-orbit splitting ($\Delta_{0}=\Gamma_{5\mathrm{v}} -\Gamma_{4\mathrm{v}}$) and the direct band-gap (E$_0=\Gamma_{1\mathrm{c}} -\Gamma_{5\mathrm{v}}$) at $\Gamma$ is 0.22 eV and 1.13 eV respectively E$_\mathrm{F}$ is the Fermi energy and the top of the valence band is set to zero energy scale.}
  \label{fig:fgr3}
\end{figure*}
It should be noted that there is no crystal-field splitting in case of zincblende structures. In order to further resolve the SOC and crystal-field splitting parameters, a quasi-cubic model of Hopfield \cite{Hopfield1960} is often used. However, due to the parametric symmetry in the equation, a thorough knowledge of the transition intensities are also required \cite{Dingle1971}. Such analyses are altogether a different methodology and thus remain beyond the scope of this work. Instead, we outline below the splitting energy differences from our analysis, both in the presence and absence of SOC in ML GaAs.\\
The effect of crystal-field splitting on the valence band in the absence of SOC is demonstrated in Fig. \ref{fig:fgr3} (a-c) by projecting the partial density-of-states (p-DOS) of As atom on the energy eigenvalues along the BZ route. We observe in Figs. \ref{fig:fgr3} (a) and (b) that at the $\Gamma$ point, the contribution from the p$_{x}$ and p$_{y}$ orbitals forms a doubly-degenerate highest valence state $\Gamma_{3\mathrm{v}}$. The next highest valence state $\Gamma_{1\mathrm{v}}$ is formed by the singly degenerate As p$_{z}$ orbital (Fig. \ref{fig:fgr3} (c)). The difference $\Gamma_{3\mathrm{v}}$-$\Gamma_{1\mathrm{v}}$= -1.95 eV, signifying a strong crystal-field splitting. In the presence of SOC, $\Gamma_{3\mathrm{v}}$ splits to Kramers pair $\Gamma_{5\mathrm{v}}$, $\Gamma_{6\mathrm{v}}$ with orbital weight from As 4P$_{3/2}$, total angular momentum, m$_\mathrm{j}$=$\pm$3/2 . The state $\Gamma_{4\mathrm{v}}$  has contribution from As 4P$_{3/2}$, m$_\mathrm{j}$=$\pm$1/2 (Fig. \ref{fig:fgr3} (e-f)). We also find that at $\Gamma$, the orbital weight of the state $\Gamma_{1\mathrm{c}}$ comes from Ga 3S$_{1/2}$ m$_\mathrm{j}$=$\pm$1/2, while $\Gamma_{4\mathrm{v}}^{\mathrm{cf}}$ has major contribution from As 4P$_{3/2}$ m$_\mathrm{j}$=+1/2.  Thus the states $\Gamma_{1\mathrm{c}}$, the doublet $\Gamma_{5\mathrm{v}}$ and $\Gamma_{6\mathrm{v}}$; and $\Gamma_{4\mathrm{v}}$ together appear to be like a sp$^2$-sp$^3$ hybridized state (see Fig. S10 in the Supplemental Material \cite{Supplemental} for details of individual atomic p-DOS). The valence spin-orbit splitting ($\Delta_{0}=\Gamma_{5\mathrm{v}} -\Gamma_{4\mathrm{v}}$) and the direct band-gap (E$_0=\Gamma_{1\mathrm{c}} -\Gamma_{5\mathrm{v}}$) at $\Gamma$ is 0.22 and 1.13 eV respectively. The presence of SOC makes $\Gamma_{1\mathrm{v}}$-$\Gamma_{4\mathrm{v}}^{\mathrm{cf}}$= -0.11 eV.\\
The effect of quasi-particle (QP) corrections is shown in Fig. \ref{fig:fgr4} (a) where the use of a G$_0$W$_0$ formalism estimate the self-energy correction of 1.82 eV to E$_0$ leading to a gap of 2.95 eV. The corresponding valence spin-orbit splitting improvements are insignificant. Here we note that, the most favourable bonding structure for group III elements is sp$^2$ hybridization which results in a trigonal planar structure. In contrast, due to higher ionic radius in group V elements; P, As and Sb prefers trigonal pyramidal structure with sp$^3$ bonding.  Therefore, a stable ML GaAs is expected to showcase a mixture of sp$^2$-sp$^3$ hybridization which results in a prominent buckling \cite{Houlong2013}. As shown in Fig. \ref{fig:fgr4} (b), we verified the thermodynamic stability from the phonon dispersion, which portrays no soft-modes. At $\Gamma$ of the BZ, all the optical out-of-plane ZO (159 cm$^{-1}$), and the degenerate in-plane longitudinal (LO) and transverse (TO) modes (264 cm$^{-1}$) are infra-red as well as Raman active.\\
\begin{figure*}[ht]
\includegraphics[width=1.6\columnwidth]{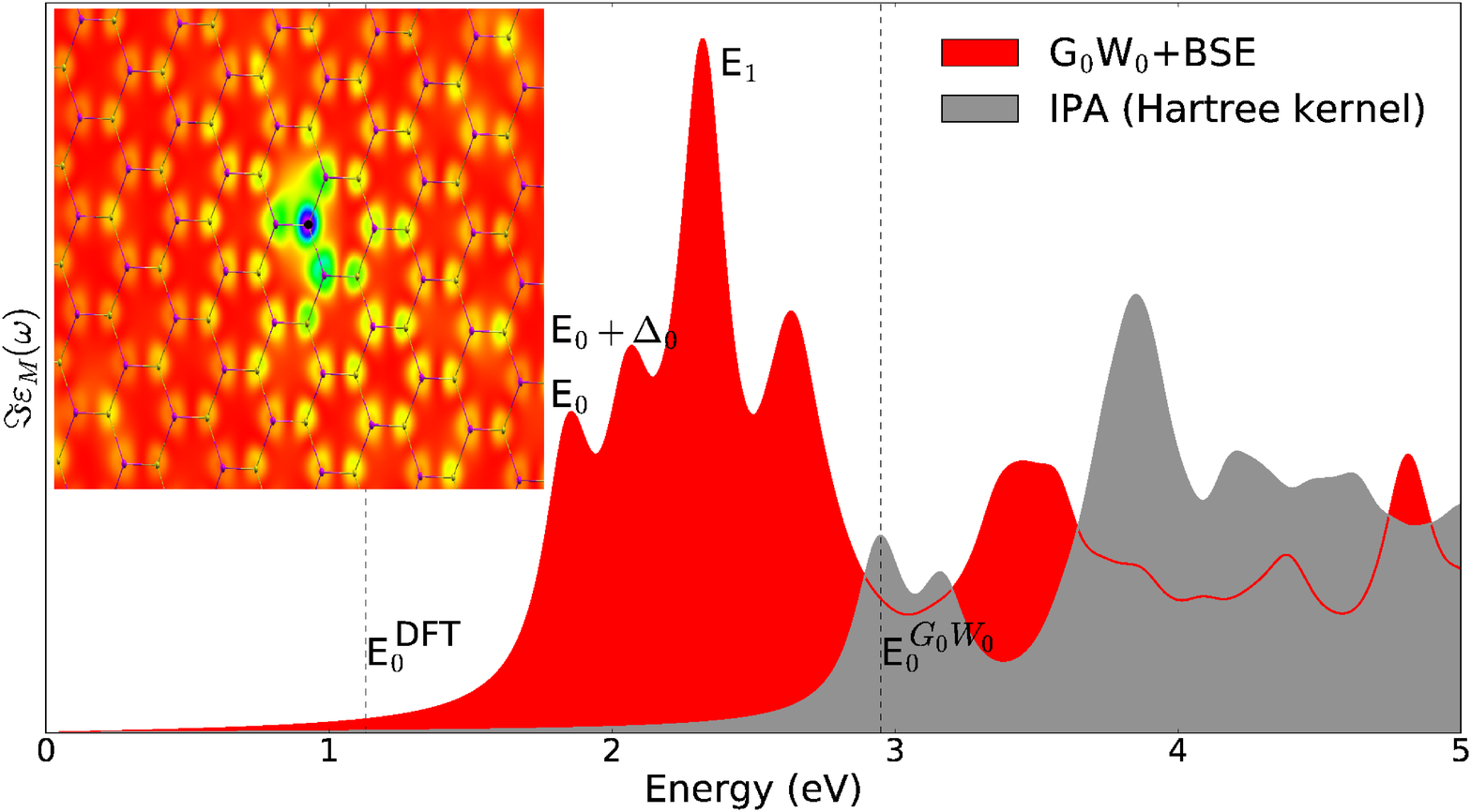}
\caption{Absorption spectra (i.e., the imaginary part of the macroscopic dielectric function $\varepsilon_{M}\left(\omega\right)$) as function of photon energy using G$_0$W$_0$+BSE and IPA (Hartree-kernel) energy corrections. The dotted lines are the DFT and G$_0$W$_0$ gap energies at 1.13 and 2.95 eV respectively. The inset shows exciton wave-function unfolded over the ML GaAs real space lattice. To know about the exciton wave-function radius, a hole was kept on the top of As atom at a distance of 1 $\mathring{\mathtt{A}}$. This is indicated by a dark spot in the inset.}
  \label{fig:fgr5}
\end{figure*}
The absorption spectra are obtained by solving a coupled electron-hole Green's propagator with the screen-exchange (SEX) potential in the self-energy (see Appendix Eqns. (\ref{W_scre}), (\ref{V_scre}) and (\ref{BSE_3})). The excited state energy corrections were borrowed from the preceding G$_0$W$_0$ calculations. The spectra in Fig. \ref{fig:fgr5} demonstrate three relevant peaks in the energy range of 1.8-2.5 eV. These peaks correspond to three inter-band critical points \cite{Lautenschlager1987} E$_0$, E$_0$+$\Delta_0$ and E$_1$ (see Fig. \ref{fig:fgr4} (a)) with exciton peak positions at 1.85 eV, 2.07 eV and 2.32 eV respectively. E$_0$ and E$_1$ critical points are due to the transition from highest valence band to the lowest conduction band at $\Gamma$ and $\textbf{K}$ points of the BZ respectively. E$_0$+$\Delta_0$ corresponds to transition from the next highest valence band to the lowest conduction band at $\Gamma$. The first two excitonic peaks are due to the spin-splitting of the valence band in which we see that the peak position difference is exactly $\Delta_0$ in the electronic dispersion. From the exciton peak positions, we measure a binding energy of 1.10 eV for the lowest exciton. This first peak E$_0$ is formed by a pair of degenerate dark and bright exciton. We find here that intra-valley transitions with anti-parallel spins produce bright excitons and vice versa. This showcases that the excitons forming from E$_0$+$\Delta_0$ as well as E$_1$ (the bottom conduction and top valence band at \textbf{K} is formed by Ga 3S$_{1/2}$, m$_\mathrm{j}$=-1/2 and As 4P$_{3/2}$, m$_\mathrm{j}$=+3/2 respectively) transitions are both bright and non-degenerate. In the non-interacting picture upto the level of Hartree kernel, the absorption spectra blue-shifts entirely with reduced peak amplitudes and broadened valleys. Interestingly, the E$_0$ exciton wave function unfolds over the real space ML lattice with a radius of approximately 16 $\mathring{\mathtt{A}}$ ($\sim$ four unit cells). Such large spanning characterizes it to be a Frenkel exciton \cite{Pope1999}.
\begin{figure*}[!ht]
\includegraphics[width=1.8\columnwidth]{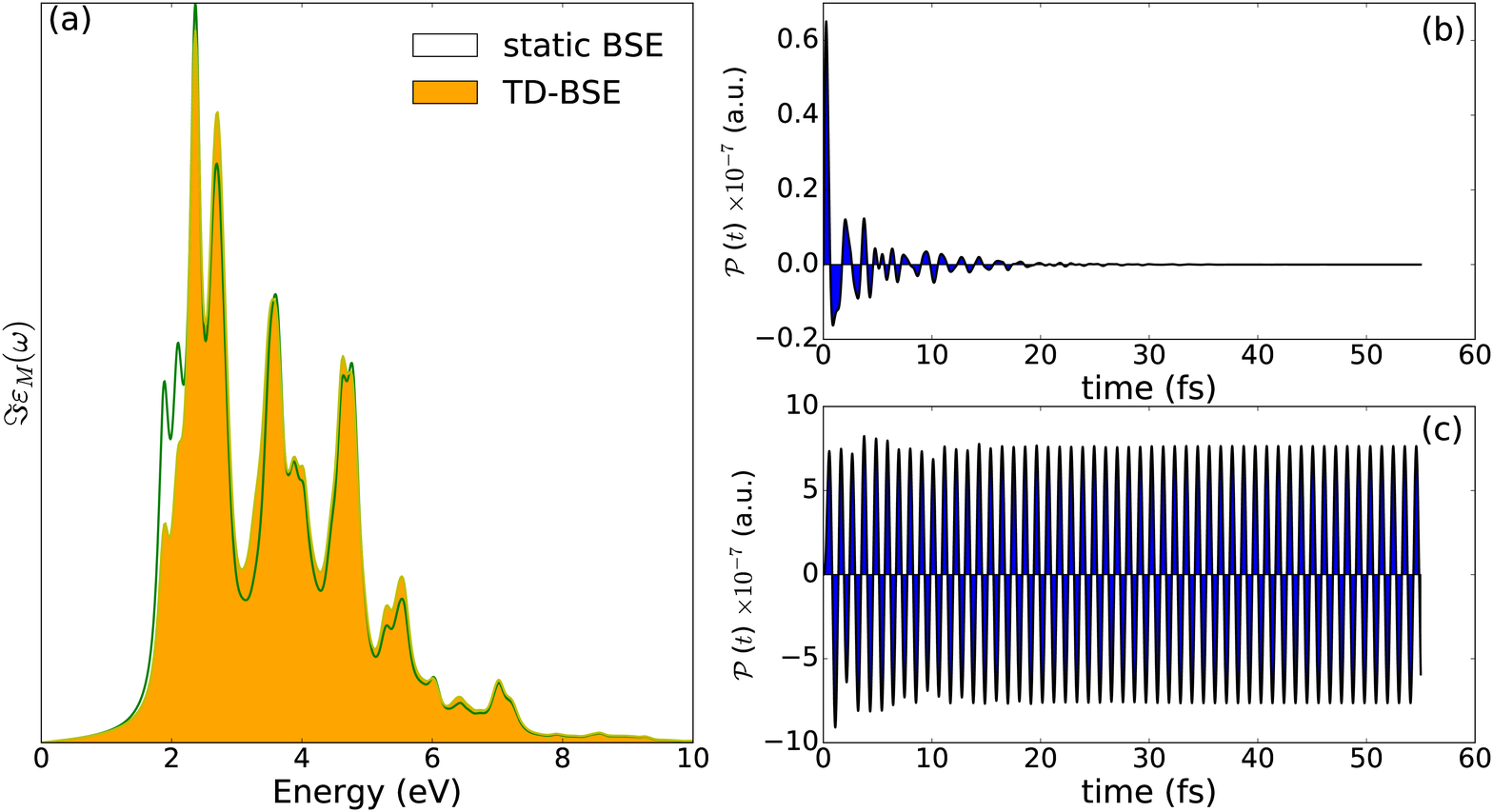}
\caption{(a) Comparisons between the absorption spectra of ML buckled GaAs calculated with real-time approach (see Appendix Eqns. (\ref{NL_1}) and (\ref{TD_BSE})) and direct (static) BSE (see Appendix Eqn. (\ref{BSE_5})). A scissor is added in both the cases to mimic G$_{0}$W$_{0}$ corrections. The real-time spectra is obtained by applying a delta-like electric field of intensity 500 KWcm$^{-2}$. The corresponding induced electric polarization is shown in (b). (c) Induced polarization along crystal $b$-axis in the presence of a quasi-monochromatic electric field along the $a$-axis. The initial sudden response can be seen in which vanishes almost at 32 fs.}
  \label{fig:fgr6}
\end{figure*}
\begin{figure*}[!ht]
\includegraphics[width=1.5\columnwidth]{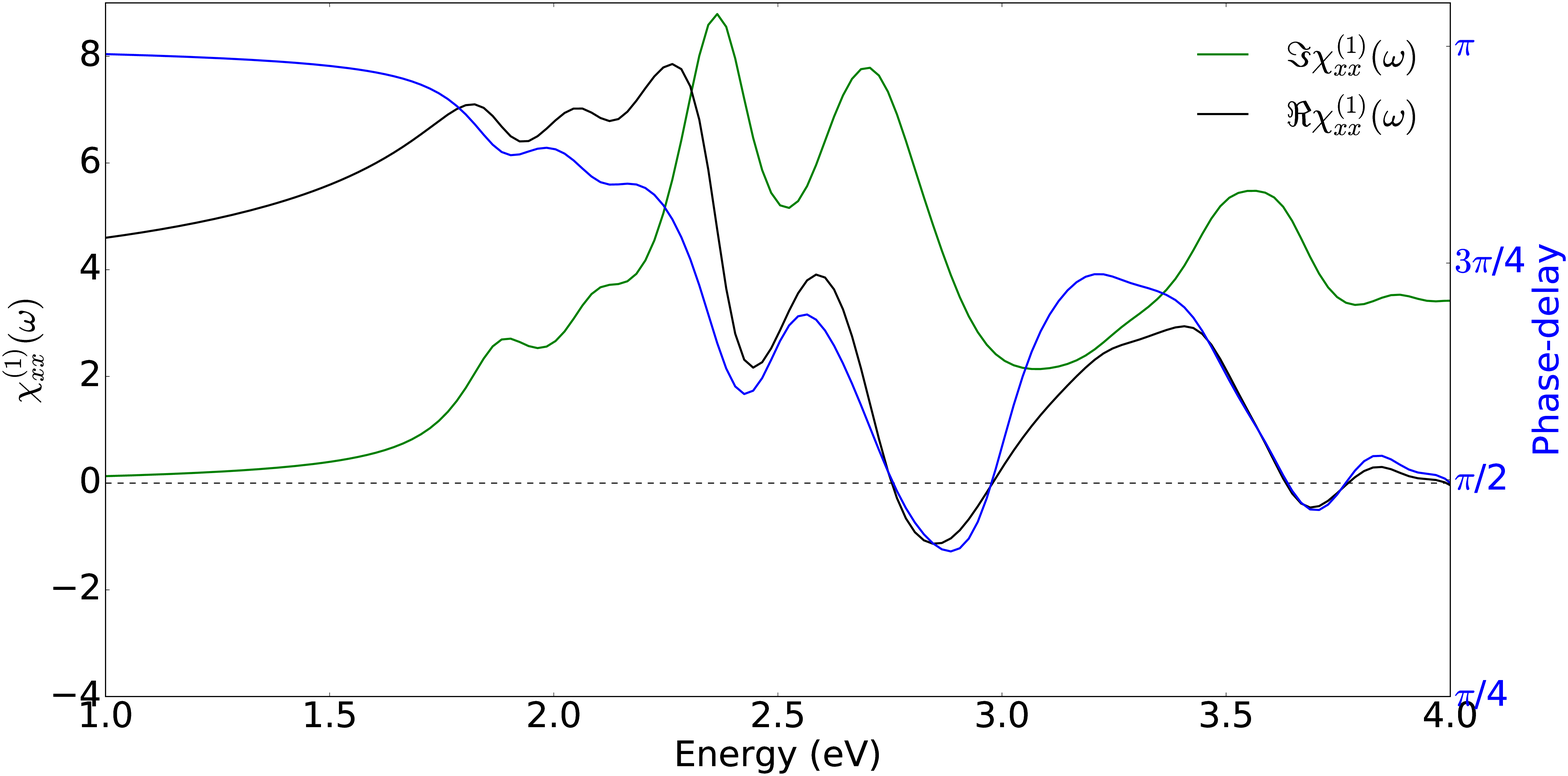}
\caption{$\Re\chi^{(1)}_{xx}(\omega)$ and $\Im\chi^{(1)}_{xx}(\omega)$ as function of the applied frequency. The responses are evaluated using the real-time approach within the TD-BSE level of approximation. The corresponding phase delay about $\pi/2$ between the electric polarization and the applied electric field as function of the applied frequency is also shown.}
  \label{fig:phase}
\end{figure*}
\subsection{Nonlinear Response}
In order to capture the NL responses, we solve in real-time the time-dependent Schr{\"o}dinger equation in presence of an external electric field $\mathcal{E}(t)$. The induced macroscopic polarization is calculated using King-Smith and Vanderbilt formalism \cite{King1993} which uses a Berry's geometric phase of the momentum state \textbf{k} (see the Appendix Eqn. (\ref{Pol})). The symmetry group arguments in buckled ML GaAs leads to non-vanishing susceptibility tensor elements \cite{Boyd2008} $\chi_{bbb}^{(2)}$=-$\chi_{baa}^{(2)}$=-$\chi_{aab}^{(2)}$=-$\chi_{aba}^{(2)}$ in which $a$ and $b$ are the in-plane ML axes. In order to verify first that the time dependent Schr{\"o}dinger equation with the SEX kernel in the system Hamiltonian (i.e., the TD-BSE) converge to a time independent excitonic BSE at weak intensities, we hit our ML system with a delta-like electric field at low intensity of 500 KWcm$^{-2}$. $A$ $priori$ Hartree and the SEX collision integrals were carried out with an energy cut-off of 35 Ry to evaluate the Hartree+SEX self-consistent potentials. We note here that one can also calculate such potentials at each step in the dynamics, for which the results are same. The EOM is then integrated using a Crank-Nicholson algorithm \cite{Crank1947} at a time-step of 10 as. 
\begin{figure*}[!ht]
\includegraphics[width=1.78\columnwidth]{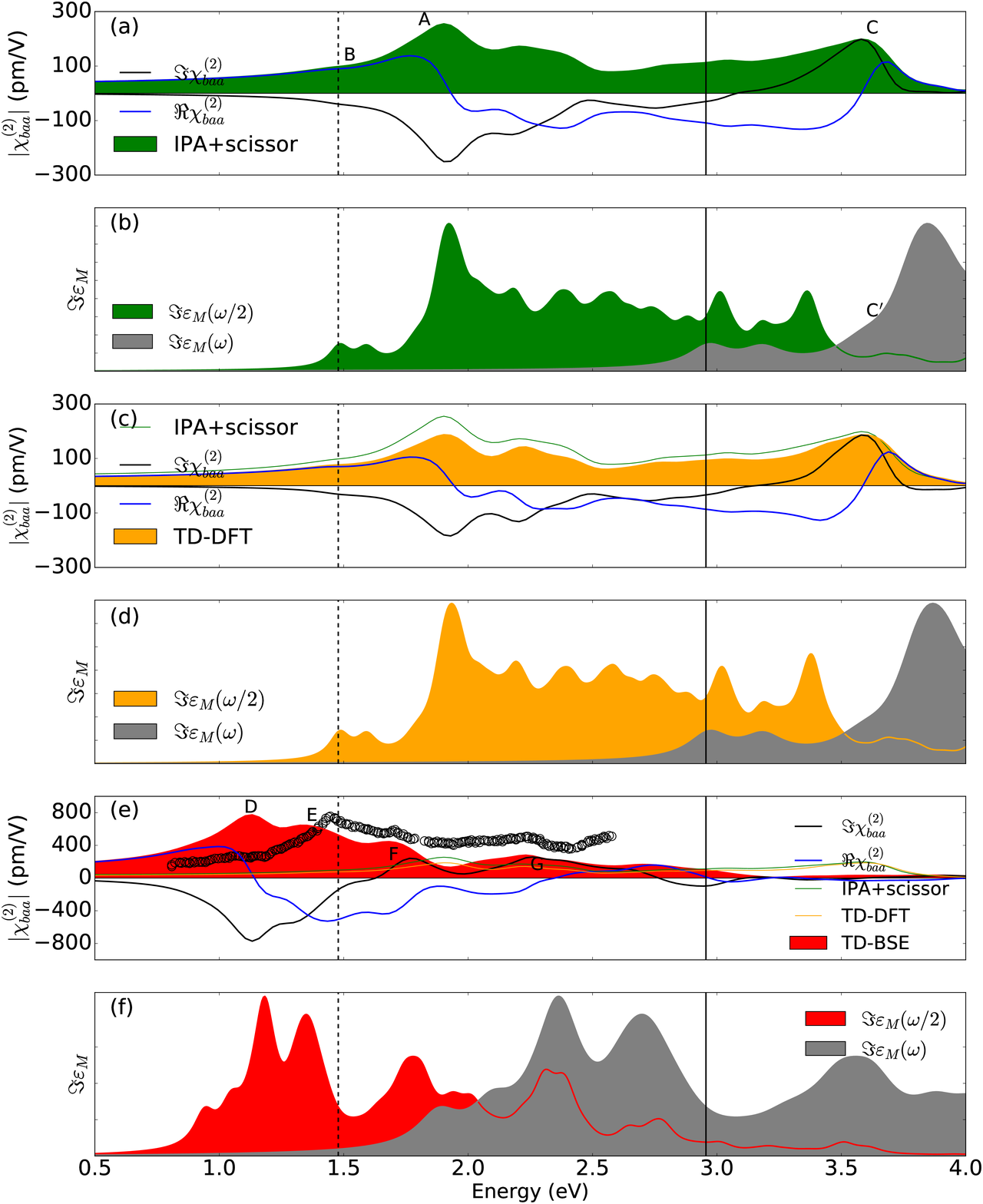}
\caption{Nonlinear SHG spectra of ML buckled GaAs as function of laser frequencies. (a), (c) and (e) are the SHG computed using IPA+scissor, TD-DFT and TD-BSE level of theory respectively (all are respective absolute values). (b), (d) and (f) shows the absorption spectra computed at $\omega$ and $\omega/2$ under the IPA+scissor, TD-DFT and TD-BSE level of theory respectively. The open circles are the experimental SHG spectra of zincblende bulk GaAs taken from Bergfeld and Daum \cite{Bergfeld2003} and put here for comparison with our monolayer structure. The solid and dashed vertical lines are for the $\omega$ and $\omega/2$ gaps. The imaginary and real parts for each of the theory are also presented. From the imaginary part, one can see that $\Im\chi_{baa}^{(2)}$ goes to zero below half of the band-gap in all the cases. All of these computations were performed on 72$\times$72$\times$1 $k$-point grid.}
  \label{fig:fgr7}
\end{figure*}
\begin{figure*}[ht]
\includegraphics[width=1.78\columnwidth]{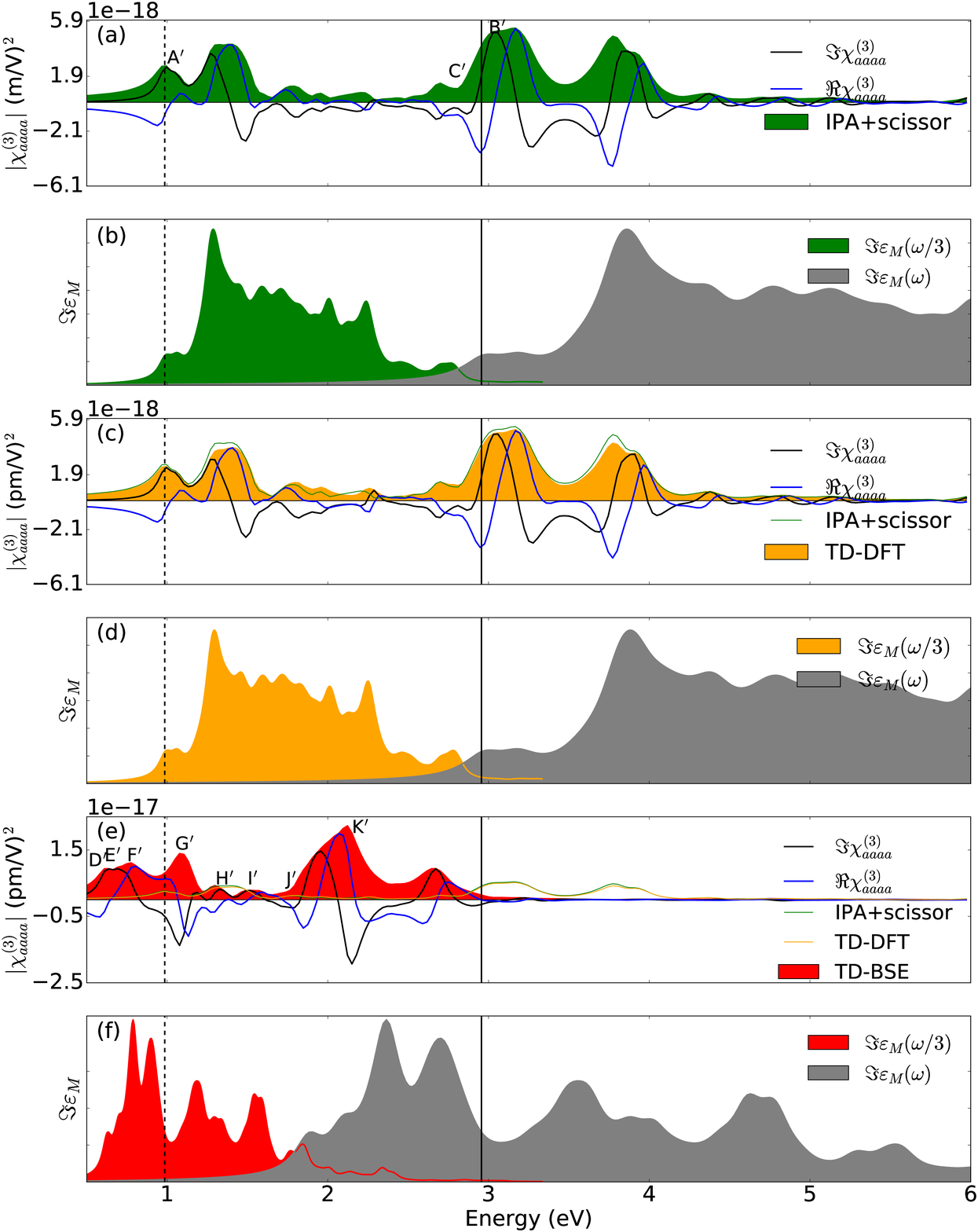}
\caption{Nonlinear THG spectra of ML buckled GaAs as function of laser frequencies. (a), (c) and (e) are the THG computed using IPA+scissor, TD-DFT and TD-BSE level of theory respectively (all are respective absolute values). (b), (d) and (f) shows the absorption spectra computed at $\omega$ and $\omega/3$ under the IPA+scissor, TD-DFT and TD-BSE level of theory respectively. The solid and dashed vertical lines are for the $\omega$ and $\omega/3$ gaps. The imaginary and real parts for each of the theory are also presented. From the imaginary part, one can see that $\Im\chi_{aaaa}^{(3)}$ goes to zero below one-third of the band-gap in all the cases. All of these computations were performed on 72$\times$72$\times$1 $k$-point grid.}
  \label{fig:fgr8}
\end{figure*}
The application of a delta-like impulse probes the system response at all frequencies. The linear $\chi_{ij}^{(1)}$ is $a$-$posteriori$ calculation and we find to be in excellent agreement to the response obtained directly by solving the time-independent BSE. This is demonstrated in Fig. \ref{fig:fgr6} (a)-(b). It should be noted that a sudden switching-on of the external field sparks spurious initial nonlinearities in the macroscopic polarization function, which should be filtered out (see Fig. \ref{fig:fgr6} (c)). For this, a finite dephasing time of about 7 fs is added in the EOM Hamiltonian amounting to a damping of 0.2 eV. This phenomenological damping \cite{Claudio2011} also mimics the presence of any experimental dissipative effects like electron scatterings, defects and lattice vibrations. We run this simulation for 55 fs. We carefully note that beyond the dephasing time of about 32 fs (almost 5 dephasing time constants), all the spurious components in the polarization are exponentially diminished. The extraction of the responses are evaluated in this time window between 32 fs to 55 fs. Figure \ref{fig:phase} demonstrate the linear responses $\Re\chi^{(1)}_{xx}(\omega)$ and $\Im\chi^{(1)} _{xx} (\omega)$ extracted from the preceding discussion. The phase delay $\varphi$ is evaluated from the complex phasor $\chi^{\left(1\right)} _{xx}\left(\omega\right)=\left|\chi^{\left(1\right)} _{xx}\left(\omega\right)\right|e^{i\varphi}$. Note that the delay closely follows the $\Re\chi^{(1)} _{xx} (\omega)$. A positive delay indicates that the polarization is leading the field and vice-versa. The horizontal line at $\varphi=\pi/2$ signifies an in-plane oscillation of the polarization current $-\frac{\partial \mathcal{P}_x}{\partial t}$ with $\mathcal{E}(t)$ \cite{Myrta2016}. Plasmonic oscillations are seen when $\Im\chi^{\left(1\right)} _{xx}\left(\omega\right)\rightarrow0$, for which $\varphi\rightarrow\pi$. For non-negligible $\Im\chi^{\left(1\right)} _{xx}\left(\omega\right)$ and $\varphi>\pi/2$, the oscillations can be due to the delocalized excitons, whereas for $\varphi<\pi/2$, it could be due to the localized excitons.\\
We now demonstrate the SHG response using a quasi-sinusoidal monochromatic field for three hierarchical perturbed Hamiltonian models in the EOM namely, TD-IPA, TD-DFT and TD-BSE (see Eqns. (\ref{DFT_GW})-(\ref{TD_BSE}) in the Appendix). The external field is along $a$-axis while the SHG is computed along $b$-axis.We note here that the NL calculations are treated with a proper choice of gauge. The response functions $\chi_{ijk}^{(2)}\left(-2\omega;\omega,\omega\right)$ and $\chi_{ijkl}^{(3)}\left(-3\omega;\omega,\omega,\omega\right)$ suffers from unphysical divergences of the form $\frac{1}{\omega_{2}}$, $\frac{1}{\omega_{2}^{2}}$ and $\frac{1}{\omega_{3}}$, $\frac{1}{\omega_{3}^{2}}$ appearing as leading terms in the respective responses. While indeed Aevrsa et. $al.$ \cite{Aversa1995} used a length gauge, they do not suffer from such unphysical divergences. In fact, it is in the velocity gauge that these divergences occur for $\omega\rightarrow$0, unless removed by using separation of intra and interband processes and/or using sum rules to eliminate them. The length gauge formalism, popular for molecules, has difficulties for an infinite periodic solid in formulating the matrix elements. The problem of unphysical divergences using different gauges was thoroughly addressed in a series by Sipe \cite{ Sipe1990, Sipe1996, Sipe2000, Sipe2007}, both with Aversa \cite{Aversa1995} and earlier with Ghahramani \cite{Sipe1993} and based on a careful separation of intra and interband contributions to the matrix elements all these divergences can be avoided. Such divergences were overcome in Sipe's work \cite{Sipe1993} and the consistency of both gauge choices was explicitly demonstrated. The sum over band transition approach within the IPA (whether formulated in length or velocity gauge) gives expressions for NLO susceptibilities that show resonances both at $\omega$ and 2$\omega$ (and in third order also at 3$\omega$).  This gives a more complete picture of the interplay of such different terms to the $\chi_{ijk}^{(2)}\left(-2\omega;\omega,\omega\right)$ or $\chi_{ijkl}^{(3)}\left(-3\omega;\omega,\omega,\omega\right)$. Unfortunately, this is only possible within the long-wavelength limit IPA, and the present formulation which includes local field contributions and even excitonic effects, does not allow for as clear a separation of these different types of contributions to the NLO response. Therefore, one resorts to a qualitative comparison to linear response at 1/2 and 1/3 of the frequency. We therefore use this approach to distinguish the 1-, 2- and 3$\omega$ resonances by comparing the spectra of $\left|\chi_{baa}^{(2)}\left(\omega\right)\right|$ and $\left|\chi_{aaaa}^{(3)}\left(\omega\right)\right|$ with the $\Im\varepsilon_{M}\left(\omega\right)$ and $\Im\varepsilon_{M}\left(\omega/2\right)$; and $\Im\varepsilon_{M}\left(\omega\right)$ and $\Im\varepsilon_{M}\left(\omega/3\right)$ for each of the IPA, TD-DFT and TD-BSE case respectively. These comparisons to linear response are possibly useful but just a qualitative comparison and are customarily used by the scientific community \cite{Takimoto2007, Myrta2014, Claudio2017, Rao2017, Hu2017, Claudio2019}. Figure \ref{fig:fgr7}(a) exhibits the SHG response for IPA. To this IPA and for all subsequent discussions, a rigid shift in the gap has been opened by a scissor operator, which is sufficient instead of the expensive G$_0$W$_0$ correction since we target to achieve the information of 1, 2 or 3$\omega$ resonances that influence the response. These processes in SHG is extracted by comparing with $\Im\varepsilon_{M}\left(\omega\right)$ and $\Im\varepsilon_{M}\left(\omega/2\right)$, at the same Hamiltonian level as shown in Fig. \ref{fig:fgr7} (b). As $\Im\chi_{ij}^{(1)}$ is the linear response, a delta-like field as mentioned in the preceding paragraph is then required for each of the case. The SHG IPA spectra are characterized by a number of peaks in the energy range 1.7-3.8 eV. The most prominent peak (A) is at 1.91 eV. There exists also a small shoulder peak B near 1.46 eV. When compared to $\Im\varepsilon_{M}\left(\omega/2\right)$, this shoulder peak is found due to the E$_0$ electronic transition at $\Gamma$ and is a 2$\omega$ resonance. The peak A is due to the E$_1$ transition at \textbf{K} and is also a 2$\omega$ resonance. Apparently, the peak near 3.65 eV (C) corresponding to a small hump (C$^{\prime}$) in the $\Im\varepsilon_{M}\left(\omega\right)$, is mainly due to a 1$\omega$ resonance. The modifications in the NL and linear spectra in TD-DFT case (Figs. \ref{fig:fgr7} (c)-(d)) are due to the presence of time-dependent Hartree and exchange-correlation potentials respectively. In the presence of the SEX kernel, the TD-BSE spectra modifies remarkably as shown in Fig. \ref{fig:fgr7} (e)-(f). This is due to the incorporation of exciton dynamics which is absent in the previous two cases. The shoulder peak B as detected in IPA is now completely smeared out and a prominent peak D at 1.16 eV emerges due to the $E_{1}$ transition. This is due to a 2$\omega$ resonance for which $\left|\chi_{baa}^{(2)}\right|$=780 pm/V and almost two to five times larger than the spectra of exfoliated ML MoS$_2$ when computed with the ``sheet" model \cite{Clark2014, Yilei2013}. Similarly, the second and the third peaks, E and F respectively, are also due to a 2$\omega$ resonance. Unlike the other cases, the fourth peak G is both due to 1- and 2$\omega$ resonances. In all the above three situations, the main contribution to the 2$\omega$ resonance comes from E$_{1}$ transition (see Fig. \ref{fig:fgr4}(a)) which means that this optical phenomenon is a \textbf{K}-point transition instead of $\Gamma$ transition. This observation is based on the fact that the E$_1$ transition in linear optics already gives apparently a stronger oscillator strength (peak height in the $\Im\chi^{2}_{baa}$) in the exciton than the E$_0$ or E$_0$+$\Delta_{0}$ transitions. The same holds true for the 1$\omega$ resonance as well in this case. Therefore, we can conclude that the second harmonic generation in buckled ML GaAs is indeed a \textbf{K} point phenomena having minor contributions from $\Gamma$ transitions. In order to calculate the static response $\chi_{baa}^{\left(2\right)}\left(\omega=0\right)$, one need to put a zero dephasing rate. However, this would mean an infinite simulation time. Instead, we extrapolate $\left|\chi_{baa}^{(2)}\left(\omega\right)\right|$ to obtain $\left|\chi_{baa}^{\left(2\right)}\left(\omega=0\right)\right|$=150 pm/V.\\
Similar processes are also seen in case of monolayers of hexagonal BN, MoS$_2$ \cite{Myrta2014} and as well as in zincblende bulk III-V compounds \cite{Myrta2016}. In Fig. \ref{fig:fgr7}(e) we have particularly shown the experimental SHG of zincblende bulk GaAs from Bergfeld and Daum \cite{Bergfeld2003}. One can observe that the transparency region is below 1 eV where only 2$\omega$ resonances contribute with a peak of 750 $\pm40$ pm/V at nearly 1.5 eV.  \\
In case of THG, we keep all the NL parameters same as in SHG, except the intensity which is increased to 10$^6$ KWcm$^{-2}$ to achieve a significant response. The external field is along $a$-axis and thus we look for $\left|\chi_{aaaa}^{(3)}\left(\omega\right)\right|$. 
In a similar approach to SHG, we start with IPA as shown in Fig. \ref{fig:fgr8} (a) and compare the respective spectra with $\Im\varepsilon_{M}\left(\omega\right)$ and $\Im\varepsilon_{M}\left(\omega/3\right)$ in Fig. \ref{fig:fgr8} (b). The first peak A$^{\prime}$ is in the transparency region at 0.98 eV, is due to the E$_0$ transition and is a 3$\omega$ resonance. The most prominent peak B$^{\prime}$ at 3.16 eV is due to E$_0$+$\Delta_0$ transition and is a 1$\omega$ resonance. The on-set of this peak near 2.81 eV (C$^{\prime}$) is however due to both 1- and 3$\omega$ resonance. Similar to SHG, the THG response variations in TD-DFT (Fig. \ref{fig:fgr8} (c) and (d)) is because of time-dependent Hartree and exchange-correlation potentials respectively. However, the SEX kernel modifies the response in Fig. \ref{fig:fgr8} (e) remarkably. Red-shifted peaks in the transparency region at 0.62 (D$^{\prime}$) and 0.69 eV (E$^{\prime}$) emerge due to spin-splitted $E_{0}$ and $E_{0}$+$\Delta_{0}$ transitions respectively (in-fact the difference when multiplied by 3 gives $\Delta_{0}$) and is a 3$\omega$ resonance. Likewise, the peak F$^{\prime}$ at 0.78 eV is again a 3$\omega$ resonance due to $E_{1}$ transition. The peak G$^{\prime}$ near 1.08 eV is the most prominent 3$\omega$ resonance in the same region and emerges due to a broadened shoulder peak around the same energy as shown in Fig. \ref{fig:fgr8} (f). At this energy, we note the $\left|\chi_{aaaa}^{(3)}\right|=$1.4$\times$10$^{-17}$ m$^{2}$/V$^{2}$. This is in two order magnitude larger than reportedly exfoliated ML MoS$_2$ and WSe$_2$ \cite{Wang2014, Henrique2018}. The next two consecutive humps H$^{\prime}$ and I$^{\prime}$ are again due to a 3$\omega$ resonance. The giant peak K$^{\prime}$ at 2.13 eV is a 3$\omega$ resonance and is due to the $E_1$ transition. The on-set of this peak near 1.76 eV (J$^{\prime}$) is due to both 1- and 3$\omega$ resonance. Hence, we can conclude that major contributions to the 3$\omega$ resonance are coming from E$_{0}$, E$_0$+$\Delta_0$ and E$_{1}$ which makes it $\Gamma$ as well as \textbf{K} point optical phenomena. However, 3$\omega$ resonance in this case is entirely coming from E$_{1}$ transition i.e. a \textbf{K} point phenomena and the most prominent one in the entire THG spectra. In addition, both the Figs. \ref{fig:fgr7} and \ref{fig:fgr8} also show the real and imaginary parts of the respective responses. One can clearly see that $\Im\chi_{baa}^{(2)}$ and $\Im\chi_{aaaa}^{(3)}$ goes to zero below half and one-third of the band-gap in all the respective cases. The two parts in each case are related through the well-known Kramers-Kronig relations with the exception of self-induced changes ($\Im\chi_{aaaa}^{(3)}(\omega; \omega, \omega, -\omega)$) in the refractive index \cite{Boyd2008}.\\
One should note that the above interesting observations are based only on a qualitative comparison to the linear spectra at $\omega$/2 and $\omega$/3 to identify the 2$\omega$ and 3$\omega$ resonances. Moreover, the E$_0$ and E$_0$+$\Delta_0$ related peaks are seem to be more prominent in linear and THG spectra but not in SHG. In order to understand such smearing-out in presence of excitons, the sum over bands approach by Sipe $et$ $al.$ \cite{Aversa1995} might give additional insights into this by the possibility to examine the matrix elements. On the other hand, one then might miss the specific excitonic enhancement effects. At this point a more theoretical development is needed to understand this problem of smearing out of the spin-peaks in SHG, which for now can be attributed to a mere observation only. A similar smearing situation is also seen in case of various other TMDCs \cite{Hua2017}. \\ 
In addition, optical signals are quite prone to the choice of substrate. But, typically dielectric coated substrates are best choice as it reduces the bulk SHG, although there could be some contribution also coming from the surface SHG. Moreover surface roughness in ML (for eg, in 2D tetragonal structures) may lead to large charge inhomogeneities which may also react with the underlying substrate. Interface defects are great problematic leading to trap and charge transference and can significantly modify the SHG and THG signals. There are reports that metal substrates fits good for hexagonal structures of TMDCs, for example CVD thin gold for 2D MoS$_{2}$ \cite{Zeng2015}, whereas Si/SiO$_2$ are good for THG signals from MoS$_{2}$ \cite{Wang2014}. For 2D TMMCs, CVD Si/SiO$_{2}$ suits as a good substrate for SHG \cite{Xu2015}. Theoretically, DFT calculations show that substrate choice for tetragonal 2D III-Vs within 1$\%$ of lattice mismatch could be metallic (100) surfaces of Pt, Cu, Pd \cite{ Houlong2013}. A scope therefore opens for careful analysis of substrate choice for wurtzite III-V monolayers.  Finally, in all our NL computations the peak electric field is of the order 10$^7$ Vm$^{-1}$. Such values are in-fact weak enough to initiate a Zener breakdown.
\section{Conclusions}\label{conclu}
To summarize, we use the first principles many-body approach and evaluate both linear as well as nonlinear responses in buckled ML GaAs. We go beyond the static theory of the BSE to evaluate the nonlinear optical harmonic generations. The inclusion of exciton dynamics significantly enhances both the SHG and THG responses compared to ML TMDCs and TMMCs. The prominent SHG and THG responses are due to the 2- and 3$\omega$ resonance and falls in the transparency region respectively. These giant responses open the possibility for the generation of nonlinear signals from ML GaAs.
\begin{acknowledgements}
Both the authors acknowledge the computational support from Institute's Central Computational Facility, CCF. HM acknowledges MHRD, Govt. of India, for providing fellowship.
\end{acknowledgements}
\appendix*
\section{} 
\subsection{Linear spectra: G$_{0}$W$_{0}$ and Bethe-Salpeter Equation}
Within the linear response many body perturbation theory (MBPT), we first start with the electronic Hartree-Fock exchange static self-energy $\textstyle \sum_{n\textbf{k}}^{x}$. The matrix elements of this self-energy in the plane wave basis set are diagonal and can be expressed as \cite{Rohlfing2000, Fetter2013}
\begin{multline}\label{Self_Hartree}
\textstyle \sum_{n\textbf{k}}^{x}=\left\langle n\textbf{k}\left|\sum^{x}\right|n\textbf{k}\right\rangle =-\textstyle \displaystyle \sum_{m}^{occ}\int_{BZ}\frac{d\textbf{q}}{\left(2\pi\right)^{3}}\sum_{\textbf{G}}v\left(\textbf{q}+\textbf{G}\right) \\
\times\left|\rho_{nm}\left(\textbf{k},\textbf{q},\textbf{G}\right)\right|^{2}f_{m\left(\textbf{k}-\textbf{q}\right)}
\end{multline}
in which $\left|n\textbf{k}\right\rangle$ is the momentum state of $n^{th}$ band, $\textbf{q}$ are the transferred momenta, $m$ are the number of occupied electronic bands, $f$ is the Fermi function and $\rho$ is the density matrix.  $\textbf{G}$ are the $G$-vectors with $v\left(\textbf{q}+\textbf{G}\right)=\frac{4\pi}{\left|\textbf{q}+\textbf{G}\right|^{2}}$ as the three-dimensional Coulomb potential in the Fourier transformed $\textbf{q}$-plane.\\
At this level, the Hartree-Fock contribution to the ground state Kohn-Sham energy eigenvalues can be written as
\begin{equation}\label{Ener_Hartree}
E_{n\textbf{k}}^{HF}=E_{n\textbf{k}}^{DFT}+\left({\displaystyle {\textstyle \sum_{n\textbf{k}}^{x}}}-V_{n\textbf{k}}^{xc}\right)
\end{equation}
in which $V_{nk}^{xc}$ is the exchange and correlation functional at the level of local density or generalized gradient approximation.\\
The GW approximation theory is the generalization of this Hartree-Fock theory achieved by replacing the bare static screening potential $v\left(\textbf{r},\textbf{r}^{\prime} \right)$ by a dynamic screened interaction $\textrm{W}\left(\textbf{r},\textbf{r}^{\prime};\omega\right)$. We describe this scheme as follows: Using the time-ordered single-particle non-interacting Green's propagator G$_{0}$, the polarization within the random phase approximation (RPA, i.e., using Hartree-kernel) is first calculated 
\begin{equation}\label{polar}
\mathcal{P}\left(\textbf{r},\textbf{r}^{\prime\prime};\tau\right)=-i\textrm{G}_{0}\left(\textbf{r},\textbf{r}^{\prime};\tau\right)\textrm{G}_{0}\left(\textbf{r}^{\prime},\textbf{r};-\tau\right)
\end{equation}
where $\tau= t-t^{\prime}$, while $t$ and $t^{\prime}$ are the time developments in the Green's propagator. Note that Eqn.(\ref{polar}) is summed over both the occupied and unoccupied states in the Fourier transformed $\omega$-plane
\begin{multline}
\mathcal{P}\left(\textbf{r},\textbf{r}^{\prime\prime};\omega\right)=\displaystyle\sum_{i}^{occ}\displaystyle \sum_{j}^{unocc}\psi_{i}^{0}\left(\textbf{r}\right)\psi_{j}^{0\ast}\left(\textbf{r}\right)\psi_{i}^{0\ast}\left(\textbf{r}^{\prime}\right)\psi_{j}^{0}\left(\textbf{r}^{\prime}\right) \\
\times\left[\frac{1}{\omega+E_{i}^{0}-E_{j}^{0}+i\eta}-\frac{1}{\omega-E_{i}^{0}+E_{j}^{0}-i\eta}\right]
\end{multline}
where the number $\eta$ is infinitesimal real and positive number. The microscopic dielectric function is the convolution of $\mathcal{P}\left(\textbf{r},\textbf{r}^{\prime\prime};\omega\right)$ with $v\left(\textbf{r},\textbf{r}^{\prime} \right)$
\begin{equation}\label{diel}
\varepsilon\left(\textbf{r},\textbf{r}^{\prime};\omega\right)=\delta\left(\textbf{r}-\textbf{r}^{\prime}\right)-\int \mathcal{P}\left(\textbf{r},\textbf{r}^{\prime\prime};\omega\right) v\left(\textbf{r},\textbf{r}^{\prime\prime}\right)d^{3}\textbf{r}^{\prime\prime}
\end{equation}
From Eqn. (\ref{diel}), the inverse microscopic dielectric function $\varepsilon^{-1}\left(\textbf{r},\textbf{r}^{\prime\prime};\omega\right)$ is obtained and is again convoluted with $v\left(\textbf{r},\textbf{r}^{\prime} \right)$ to get
\begin{equation}\label{dyn_W}
\textrm{W}\left(\textbf{r},\textbf{r}^{\prime};\omega\right)=\int\varepsilon^{-1}\left(\textbf{r},\textbf{r}^{\prime\prime};\omega\right)v\left(\textbf{r},\textbf{r}^{\prime\prime}\right)d^{3}\textbf{r}^{\prime\prime}
\end{equation}
Equation (\ref{dyn_W}) signifies that a quasi-particle at $\textbf{r}$ induces an effective screened interacting dynamic potential $\textrm{W}\left(\textbf{r},\textbf{r}^{\prime};\omega\right)$ at $\textbf{r}^{\prime}$.
Once $\textrm{W}\left(\textbf{r},\textbf{r}^{\prime};\omega\right)$ is known, the GW self-energy is a final full frequency-axis convolution of non-interacting propagator G$_{0}$ with $\textrm{W}\left(\textbf{r},\textbf{r}^{\prime};\omega\right)$
\begin{multline}\label{GW}
\textstyle \sum^{\textrm{GW}}\left(\textbf{r},\textbf{r}^{\prime};\omega\right)\\
=\frac{i}{2\pi}\int_{-\infty}^{+\infty}\textrm{G}_{0}\left(\textbf{r},\textbf{r}^{\prime};\omega+\omega^{\prime}\right)\textrm{W}\left(\textbf{r},\textbf{r}^{\prime};\omega^{\prime}\right)e^{i\omega^{\prime}\eta}d\omega^{\prime}
\end{multline}
Single shot GW or G$_0$W$_0$ is the condition when the non-interacting Green's function is used and the screened interaction W is only once iterated through RPA. Note that now, the screening W implicitly defines $\varepsilon_{\textbf{\textrm{G}\textrm{G}}^{\prime}}^{-1}$ in Fourier transformed \textbf{q}-plane. A pure correlational G$_0$W self-energy can be extracted from Eqn. (\ref{GW}) as
\begin{multline}\label{corr}
\textstyle \sum^{c}\left(\textbf{r},\textbf{r}^{\prime};\omega\right)=\\
\frac{i}{2\pi}\int_{-\infty}^{+\infty} \textrm{G}_{0}\left(\textbf{r},\textbf{r}^{\prime};\omega+\omega^{\prime}\right)\textrm{W}^{c}\left(\textbf{r},\textbf{r}^{\prime};\omega\right)d\omega^{\prime}
\end{multline}
leaving the pure exchange term as
\begin{multline}\label{exch}
\textstyle \sum^{x}\left(\textbf{r},\textbf{r}^{\prime};\omega\right)\\
=\frac{i}{2\pi}\int_{-\infty}^{+\infty}\textrm{G}_{0}\left(\textbf{r},\textbf{r}^{\prime};\omega+\omega^{\prime}\right)v\left(\textbf{r},\textbf{r}^{\prime}\right)e^{i\omega^{\prime}\eta}d\omega^{\prime}
\end{multline}
in which $\textrm{W}^{c}\left(\textbf{r},\textbf{r}^{\prime};\omega\right)=\textrm{W}\left(\textbf{r},\textbf{r}^{\prime};\omega\right)-v\left(\textbf{r},\textbf{r}^{\prime}\right)$. Eqn. (\ref{exch}) can be computed analytically in both $\textbf{q}$ and $\omega$ plane leading to Eqn. (\ref{Self_Hartree}). Symbolically, the total GW self-energy is split into a respective exchange (Hartree-Fock) and correlational part as $i$G$v$+$i$G(W-$v$). Because of the presence of several poles of both G$_{0}$ and W, located infinitely close to the real-frequency axis, the above frequency integral Eqn. (\ref{GW}) becomes computationally expensive. What is then done is to replace $\varepsilon_{\textbf{GG}^{\prime}}^{-1}$ with an effective inverse dielectric function model consisting of a single pole, essentially at the plasma frequency describing the collective charge-neutral excitation. This approximation is known as the ``plasmon-pole model". There has been various such celebrated approximate models developed in the past like the Hybertsen-Louie (HL) \cite{Hybertsen1985}, Godby-Needs (GN) \cite{Godby1989}, Linden-Horsch (LH) \cite{Linden1988} and Engel-Farid (EF) \cite{Engel1993}, to name a few \cite{Larson2013}. Out of these, the first two are the most common in practice. Here, we use the GN plasmon-pole model approximation since this is found to be most stable and fits the above inverse dynamic dielectric function and the corresponding QP energies very accurately when evaluated by the complete full-frequency integral \cite{Larson2013}.\\ 
The GN plasmon-pole model replace this $\varepsilon_{\textbf{GG}^{\prime}}^{-1}$ with a single pole function of the form
\begin{multline} \label{GN_PP}
\varepsilon_{\textbf{G},\textbf{G}^{\prime}}^{-1}\left(\textbf{q},\omega\right)\sim\delta_{\textbf{G},\textbf{G}^{\prime}}+R_{\textbf{G},\textbf{G}^{\prime}}\left(\textbf{q}\right)\left\{ \frac{1}{\left[\omega-\Omega_{\textbf{G},\textbf{G}^{\prime}}\left(\textbf{q}\right)+i0^{+}\right]}\right. \\
\left.-\frac{1}{\left[\omega+\Omega_{\textbf{G},\textbf{G}^{\prime}}\left(\textbf{q}\right)+i0^{+}\right]}\right\} 
\end{multline}
The residuals $R_{\textbf{G},\textbf{G}^{\prime}}\left(\textbf{q}\right)$ and the energy $\Omega_{\textbf{G},\textbf{G}^{\prime}}\left(\textbf{q}\right)$ parameters are generally obtained by fitting after calculating the RPA inverse dielectric matrix at two given frequencies $\omega$=0 and at a user defined imaginary frequency ($i\omega^{\prime}_{p}$), in which ($\omega^{\prime}_{p}$) is typically chosen such that it should be near to the plasmon frequency ($\omega_{p}$). These two parameters are then evaluated as $R_{\textbf{G},\textbf{G}^{\prime}}\left(\textbf{q}\right) = \frac{1}{2}\varepsilon_{\textbf{G},\textbf{G}^{\prime}}^{-1}\left(\textbf{q},\omega=0\right)\Omega_{\textbf{G},\textbf{G}^{\prime}}\left(\textbf{q}\right)$ and $\Omega_{\textbf{G},\textbf{G}^{\prime}}\left(\textbf{q}\right)=\omega^{\prime}_{p}\sqrt{\frac{\varepsilon_{\textbf{G},\textbf{G}^{\prime}}^{-1}\left(\textbf{q},\omega=\omega^{\prime}_{p}\right)}{\varepsilon_{\textbf{G},\textbf{G}^{\prime}}^{-1}\left(\textbf{q},\omega=0\right)-\varepsilon_{\textbf{G},\textbf{G}^{\prime}}^{-1}\left(\textbf{q},\omega=\omega^{\prime}_{p}\right)}}$. To account for the charge inhomogeneity, a local field effect was also employed along the in-plane periodic direction using a sufficient response block size cut-off.\\  
Assuming that the difference between the QP and the mean-field energies are small, the nonlinear QP energy \cite{Rohlfing2000}
\begin{equation} \label{Total_Ener}
E_{n\textbf{k}}^{QP}=E_{n\textbf{k}}^{DFT}+\left\langle n\textbf{k}\left|\textstyle \sum^{\textrm{GW}}\left(\textbf{r},\textbf{r}^{\prime};\omega=E_{n\textbf{k}}^{QP}\right)-V^{xc}\right|n\textbf{k}\right\rangle 
\end{equation}
can be linearized \cite{Mahan2014, Rohlfing2000} by taking the first-order Taylor's series expansion around the Kohn-Sham DFT eigen-energies in order to get
\begin{multline} \label{Total_Ener_QP}
E_{n\textbf{k}}^{QP}=E_{n\textbf{k}}^{DFT}\\
+Z_{nk}\left\langle n\textbf{k}\left|\textstyle \sum^{\textrm{GW}}\left(\textbf{r},\textbf{r}^{\prime};\omega=E_{n\textbf{k}}^{DFT}\right)-V^{xc}\right|n\textbf{k}\right\rangle 
\end{multline} 
The QP lifetimes are the reciprocal of the imaginary part of $\textstyle \sum_{n\textbf{k}}^{\textrm{GW}}$=$\left\langle n\textbf{k}\left|\textstyle \sum^{\textrm{GW}}\left(\textbf{r},\textbf{r}^{\prime};\omega=E_{n\textbf{k}}^{DFT}\right)\right|n\textbf{k}\right\rangle$.
This factor
\begin{equation} \label{Renorm}
Z=\left[1-\frac{d\textstyle\sum_{n\textbf{k}}^{\textrm{GW}}}{d\omega}\right]^{-1}
\end{equation}
with 0$\leq Z_{n\textbf{k}}\leq$1 is then the QP renormalized weight factor. Values of $Z$ very close to 1 signifies a pure QP state. The corresponding spectral function
\begin{multline} \label{Spectral}
A_{n,\textbf{k}}\left(\omega\right)=\frac{1}{\pi}\times \left|\Im \textstyle \textstyle \sum_{n\textbf{k}}^{\textrm{GW}}\right|\\
\times\left[\left[\omega-E_{n\textbf{k}}^{DFT}-\left(\Re\textstyle\sum_{n\textbf{k}}^{\textrm{GW}}-V_{n\textbf{k}}^{xc}\right)\right]^{2}+\left[\Im\textstyle\sum_{n\textbf{k}}^\textrm{{GW}}\right]^{2}\right]^{-1}
\end{multline}
is Lorentzian and the spreading (full-width at half maximum, FWHM) defines the strength of the correlated interaction. A sharp spectral function defines a less correlated interaction, while a dwarf and spread defines a strong interaction.\\
One of the major challenges when dealing with 2D systems, like in our case with ML GaAs, is the finite length in one of the spatial direction. This introduces rapid variations in screening and as a result the integral quantities like exchange self-energies, BS kernel, total energy expression, etc. suffers $\textbf{q}\rightarrow$0 divergence problem due to the quasi-2D nature of Coulomb interaction. In order to compute those quantities properly, ``random integration method" emerged as the most numerically accurate methodology \cite{Olivia1998, Rozzi2006, Sangalli2019}. These divergences can be solved by the state-of-the-art computational methodologies performed on high-performance CPUs. We explain this in the spirit of \cite{Olivia1998, Rozzi2006, Sangalli2019}: The numerical evaluation of the GW self-energy (Eqn. (\ref{GW})) is a horrendous task. A fine sampling of the BZ would require an exorbitant computational cost since large grids of transferred momenta are always connected with the use of equally large grids of k points \cite{Rozzi2006}. Therefore a preferable solution is to fix certain k-points grid and the integration is then performed over the BZ by using a large random grid of points to do the q -summation. These random points are chosen in such a way to cover the whole of the BZ.
Quantitatively, rewriting the Coulombic integral as $\int d^{3}\textbf{q}\left[\frac{f\left(\textbf{q}\right)}{\left|\textbf{q}+\textbf{G}\right|^{2}}\right]$, each of this kind of term appearing in static or dynamic self-energy (in GW or BSE-only for oscillators and occupation numbers) is integrated around each $\textbf{q}$ over a small volume centered at $\textbf{q}+\textbf{G}$  whereas the rest of the integrand $f\left(\textbf{q}\right)$ remains almost constant. Computationally, this small volume could be a box for planar geometry, cylinder for one-dimensional geometry and sphere for bulk. The height of the box should be equal to the either side distance between the periodic images done while using ground state or the density functional theory task. We particularly this divergence overcome situation for the diagonal matrix elements, the case with off-diagonal matrix elements is then straight forward: The diagonal matrix element of the exchange self-energy (Eqn. (\ref{Self_Hartree})) after assuming that the integral is a smooth function of momenta, can then be written as
\begin{multline} \label{Monte}
\left\langle n\textbf{k}\left|\textstyle \sum^{x}\right|n\textbf{k}\right\rangle \\
\approx\sum_{\textbf{q}_{\textbf{i}}}\sum_{\textbf{G}}F\left(\textbf{q}_{\textbf{i}},\textbf{G}\right)\int_{small_{BZ}\left(\textbf{q}_{\textbf{i}}\right)}d^{3}\textbf{q}\frac{4\pi}{\left|\textbf{q}+\textbf{G}\right|^{2}}
\end{multline}
This integral can now be evaluated using a Montecarlo method and the procedure is known in literature as random integration method. This way we see that the $\textbf{q}_{\textbf{i}}\rightarrow0$ divergence is also resolved here since the 3D $\textbf{q}$ integration forbids this to happen. In addition, the integral pre-factor is also regular when $\textbf{q}_{\textbf{i}}\rightarrow0$. 10$^{6}$ random points were incorporated in our calculation in order to evaluate the Coulomb integrals with a $\textbf{G}$-vector cut-off of 3 Ry. The numerical integral was defined within a box-structure extending 30 $\mathring{\mathrm{A}}$ on either side of the ML GaAs. This truncated the Coulomb potential between the repeated images and a faster convergence was achieved.\\
Excitonic affairs are governed by a two-particle (electron and hole) Dyson-like equation of motion. In a ladder-approximation representation \cite{Rohlfing2000},  
\begin{multline} \label{BSE_1}
\mathcal{L}\left(12;1^{^{\prime}}2^{^{\prime}}\right)=\mathcal{L}_{0}\left(12;1^{^{\prime}}2^{^{\prime}}\right)+\\
\int d\left(3456\right)\mathcal{L}_{0}\left(14;1^{^{\prime}}3\right)K\left(35;46\right)\mathcal{L}\left(62;52^{\prime}\right)
\end{multline}
in which $\mathcal{L}\left(12;1^{^{\prime}}2^{^{\prime}}\right)$ and $\mathcal{L}_{0}\left(12;1^{^{\prime}}2^{^{\prime}}\right)$ are the interacting and non-interacting two-particle Green's propagator respectively. The variable ``(1)" (and similar others) is a short hand notation for the spatial, spin and four time (two creation and two annihilation) coordinates: $\left(1\right)\equiv\left(r_{1},\sigma_{1},t_{1}\right)$ respectively. In case of occupied ($v$) and unoccupied ($c$) states, $\mathcal{L}_{0}$ in Fourier transform plane has the form 
\begin{equation} \label{BSE_2}
\mathcal{L}_{0}^{vcv^{\prime}c^{\prime}}\left(\omega\right)=\frac{1}{\omega-\left(E_{c}^{DFT}-E_{v}^{DFT}\right)+i\eta}\delta_{cc^{\prime}}\delta_{vv^{\prime}}
\end{equation}
Note here that the four time variables are now decomposed in a single frequency in the $\omega$ plane. \\
The kernel $K_{vc\textbf{k}v^{\prime}c^{\prime}\textbf{k}^{\prime}}$ is a functional static quantity and is the sum of a bare exchange Coulomb repulsion and statically screened Coulomb attraction between the electron and hole. The latter is represented as 
\begin{multline} \label{W_scre}
\textrm{W}\left(vc\textbf{k};v^{\prime}c^{\prime}\textbf{k}^{\prime}\right)=\frac{1}{\Omega}\sum_{\textbf{G}\textbf{G}^{\prime}}v\left(\textbf{q}+\textbf{G}^{\prime}\right)\epsilon_{\textbf{G}\textbf{G}^{\prime}}^{-1}\left(\textbf{q}\right)\times \\
\left\langle v^{\prime}\textbf{k}^{\prime}\left|e^{-i\left(\textbf{q}+\textbf{G}^{\prime}\right).\textbf{r}}\right|v\textbf{k}\right\rangle \left\langle c\textbf{k}\left|e^{i\left(\textbf{q}+\textbf{G}^{\prime}\right).\textbf{r}}\right|c^{\prime}\textbf{k}^{\prime}\right\rangle \delta_{\textbf{q},\textbf{k}-\textbf{k}^{\prime}}
\end{multline}
while the former is
\begin{multline} \label{V_scre}
V\left(vc\textbf{k};v^{\prime}c^{\prime}\textbf{k}^{\prime}\right)=\frac{1}{\Omega}\sum_{\textbf{G}\neq0}v\left(\textbf{G}\right)\left\langle v^{\prime}\textbf{k}^{\prime}\left|e^{-i\textbf{G}.\textbf{r}}\right|c^{\prime}\textbf{k}^{\prime}\right\rangle\times \\ 
\left\langle c\textbf{k}\left|e^{i\textbf{G}.\textbf{r}}\right|v\textbf{k}\right\rangle
\end{multline}
where $\Omega$ in this case is the cell volume. $K$ is thus defined as $K_{vc\textbf{k};v^{\prime}c^{\prime}\textbf{k}^{\prime}}=\left\langle vc\textbf{k}\left|\textrm{W}-2V\right|v^{\prime}c^{\prime}\textbf{k}^{\prime}\right\rangle$. It is in this statically screened kernel W in which the G$_{0}$W$_{0}$ QP energies are included to get the correct transition energies. Note that in order to obtain a solvable BSE \cite{Marini2003}, W is approximated to be a static, which can be borrowed from the preceding dynamic screening calculations in G$_{0}$W$_{0}$ simply by putting $\omega$=0.\\
Assuming that the off-diagonal elements in the self-energies are small which consequently makes the total Hamiltonian to be a Hermitian and the QP states orthogonal, the exciton EOM (i.e., the BSE) becomes \cite{Rohlfing2000}
\begin{multline} \label{BSE_3}
\left(E_{c\textbf{k}}^{QP}-E_{v\textbf{k}}^{QP}\right)A_{vc\textbf{k}}^{s}+ \\
\sum_{v^{\prime}c^{\prime}\textbf{k}^{\prime}}\left\langle vc\textbf{k}\left|K_{vcv^{\prime},v^{\prime}c^{\prime}\textbf{k}^{\prime}}\right|v^{\prime}c^{\prime}\textbf{k}^{\prime}\right\rangle A_{vc\textbf{k}}^{s}=E_{X}^{S}A_{vc\textbf{k}}^{s}
\end{multline}
in which $S$ is each exciton (i.e., a pair state with a distinct principal quantum number and momentum wave-vector difference between $v$ and $c$), $E_{X}$ is the excitonic energy that is obtained by diagonalizing this Hamiltonian and $A_{vc\textbf{k}}^{s}$ is the excitonic amplitude in the electron-hole basis and contains the light polarization direction. As the momentum wave-vector difference is zero for vertical transitions, therefore excitons with such transitions (bright excitons) are only detectable. The resonant Green's propagator is then
\begin{equation} \label{BSE_4}
\mathcal{L}_{vc,v^{\prime}c^{\prime}}\left(\omega\right)=\sum_{S}\frac{A_{vc\textbf{k}^{S}} A_{v'c'\textbf{k}'}^{S\ast}}{\omega-E_{X}+i\eta}
\end{equation}
The numerator can be obtained via residue theorem and signifies the exciton oscillator strength. The macroscopic dielectric function (i.e., the absorption spectra) is thus evaluated in limit of long wavelength $\textbf{q}\rightarrow0$ \cite{Rohlfing2000}
\begin{multline} \label{BSE_5}
\varepsilon_{M}\left(\omega\right)= 1- \lim_{\textbf{q} \to  0} \left(\frac{8\pi}{|q|^2\Omega}\right)\sum_{vck}\sum_{v'c'k'}\langle v \textbf{k}- \textbf{q}|e^{-i\textbf{qr}}|c \textbf{k}\rangle\times \\ 
\langle c' \textbf{k}'|e^{i\textbf{qr}}| v' \textbf{k}' - \textbf{q}\rangle \sum_{S}\ \left (\frac{A_{vc\textbf{k}}^{S} A_{v'c'\textbf{k}'}^{S\ast}}{\omega-E_{X}+i\eta}\right)
\end{multline}
This is also the linear response function $\chi_{ij}^{\left(1\right)}\left(\omega\right)$.\\  
In order to analyse if the exciton is ``Frenkel" or ``Wannier"-type, the exciton wave-function is needed. This can be written as 
\begin{equation} \label{exce}
\left|\varPhi^{S}\left(\textbf{r}_{e},\textbf{r}_{h}\right)\right\rangle =\sum_{vc\textbf{k}}A_{vc\textbf{k}}^{S}\phi_{v\textbf{k}}\left(\textbf{r}_{e}\right)\phi_{c\textbf{k}}\left(\textbf{r}_{h}\right)
\end{equation}
in which $\textbf{r}_e$ and $\textbf{r}_h$ are the electron and hole coordinates in real-space. We note that the evaluation of this wave-function would require six-coordinates. Thus, we fix the hole position on the top of As atom and obtain the projection 
$\left|\varPhi^{S}\left[\left(0,0,0\right),\left(0,0,0\right)\right]\right|^{2}$
on the $x$-$y$ plane. This has been exhibited as inset of Fig. \ref{fig:fgr5} in the main text. 

\subsection{Nonlinear spectra: Time-dependent Schr{\"o}dinger's equation}
The time-development of occupied states $\left|\nu_{nk}\right\rangle$ can be obtained from a time-dependent Schr{\"o}dinger's equation as \cite{Claudio2013}
\begin{equation} \label{NL_1}
i\hbar\frac{d}{dt}\left|\nu_{n\textbf{k}}\right\rangle =\left[\mathcal{H}_{\textbf{k}}^{system}+i\mathcal{E}(t)\cdot\tilde{\partial_{\textbf{k}}}\right]\left|\nu_{n\textbf{k}}\right\rangle 
\end{equation}
in which $\mathcal{H}_{k}^{system}$ is the system Hamiltonian and $\mathcal{E}(t)\cdot\tilde{\partial_{\textbf{k}}}$ is the coupling of the electrons with the external field. If the system is periodic, the Born-von K{\'a}rm{\'a}n periodic boundary condition impose the operator $\tilde{\partial_{\textbf{k}}}\equiv\frac{\partial}{\partial \textbf{k}}$. The solutions to this equation are then gauge invariant under unitary transformations of the Bloch state $\left|k\right\rangle$.\\
Instead of a perturbative way (i.e., in the Fourier transformed domain), Eqn. (\ref{NL_1}) is solved directly in the real-time (RT) domain. The reasons are straightforward. A perturbative scheme is always computationally very expensive whereas in RT, the many-body effects can be efficiently added in the Hamiltonian. \\
In order to find the macroscopic polarization, we follow the modern theory of polarization by King-Smith and Vanderbilt \cite{King1993}. 
These authors argued that the Berry's phase change generated by a closed path in \textbf{k}-space correctly defines the macroscopic polarization of a periodic system. When the states $\left|\nu_{nk}\right\rangle$ are known, the in-plane macroscopic polarization along the lattice vector $a$ can then be evaluated from \cite{King1993, Claudio2013}
\begin{equation} \label{Pol}
\mathcal{P}_{\parallel}=-\frac{eg_{s}}{2\pi\Omega}\frac{\left|a\right|}{N_{\textbf{k}_{\bot}}}\sum_{\textbf{k}_{\bot}}\Im \mathrm{log}\prod_{\textbf{k}_{\parallel}}^{N_{\textbf{k}_{\parallel}}-1}\mathrm{det}\mathrm{S}\left(\textbf{k},\textbf{k}+\textbf{q}_{\parallel}\right)
\end{equation}
in which $e$ is the electronic charge, $g_{s}$ is the spin degeneracy, $\mathrm{S}\left(\textbf{k},\textbf{k}+\textbf{q}_{\parallel}\right)$ is the overlap matrix between the states $\left|\nu_{n\textbf{k}}\right\rangle$ and $\left|\nu_{m\textbf{k}+\textbf{q}_{\parallel}}\right\rangle$, $N_{\textbf{k}_{\parallel}}$ and $N_{\textbf{k}_{\bot}}$ are the respective in-plane and out-of plane $\textbf{k}$-points to the polarization direction with $q_{\parallel}=\frac{2\pi}{N_{k_{\parallel}}}$.\\
The system Hamiltonian in Eqn. (\ref{NL_1}) can now be constructed as follows:
In the independent-particle approximation, the energy eigenvalues are simply evaluated from the Kohn-Sham DFT Hamiltonian \cite{Claudio2013}
\begin{equation} \label{DFT}
\mathcal{H}_{k}^{system}=\mathcal{H}_{k}^{DFT}
\end{equation}
Next, the G$_0$W$_0$ corrections can be added to this IPA Hamiltonian as either by a scissor operator \cite{Claudio2013}
\begin{multline} \label{DFT_GW}
\mathcal{H}_{\textbf{k}}^{system}=\mathcal{H}_{k}^{DFT}+\mathcal{\bigtriangleup H}^{scissor}=\mathcal{H}_{k}^{DFT}+ \\
{\textstyle \sum_{n\textbf{k}}\bigtriangleup_{n\textbf{k}}\left|v_{n\textbf{k}}^{0}\right\rangle \left\langle v_{n\textbf{k}}^{0}\right|}
\end{multline}
or directly by $\bigtriangleup_{n\textbf{k}}=E_{n\textbf{k}}^{\textrm{G}_{0}\textrm{W}_{0}}-E_{n\textbf{k}}^{DFT}$ from an $ab$-$initio$ computation.\\
The next hierarchy is the TD-DFT, where the system Hamiltonian is \cite{Takimoto2007, Claudio2013}
\begin{equation}  \label{TD_DFT}
\mathcal{H}_{\textbf{k}}^{system}=\mathcal{H}_{\textbf{k}}^{DFT}+V_{H}\left[\triangle\rho\left(r,t\right)\right]+V_{xc}\left(r\right)\left[\triangle\rho\left(r,t\right)\right]
\end{equation}
in which $V_H$ is the self-consistent Hartree potential, and $V_{xc}$ is the exchange-correlation potential at the level of Kohn-Sham DFT, now calculated quasi-statically within LDA or GGA. These two potentials are dependent on the time-varying electronic density $\rho\left(r,t\right)$. Random phase approximation is the condition when $V_{xc}$ is neglected in the system Hamiltonian. The change $\triangle\rho\left(r,t\right)=\rho\left(r,t\right)-\rho\left(r,0\right)$ is the electronic density variation and is responsible for the local-field effects due to the inhomogeneity in the system.\\
The next level of hierarchy is the incorporation of scissor-corrected SEX interaction in the Hamiltonian. This is usually known as TD-BSE \cite{Claudio2013}
\begin{equation} \label{TD_BSE}
\mathcal{H}_{\textbf{k}}^{system}=\mathcal{H}_{\textbf{k}}^{DFT}+\mathcal{\bigtriangleup H}_{\textbf{k}}^{scissor}+V_{H}\left[\triangle\rho\left(r,t\right)\right]+{\textstyle \sum_{SEX}\left[\Delta\gamma\right]}
\end{equation} 
in which $\Delta\gamma\left(r,r^{\prime},t\right)=\gamma\left(r,r^{\prime},t\right)-\gamma\left(r,r^{\prime},0\right)$ is the density fluctuation matrix induced by the external field. The self-energy ${\textstyle \sum_{SEX}}$ is the convolution between the statically screened interaction W (see below Eqn. \ref{V_scre}) and $\Delta\gamma\left(r,r^{\prime},t\right)$.\\
The EOM, Eqn. (\ref{NL_1}), is now numerically solved for $\left|\nu_{nk}\right\rangle$ using the following algorithm developed by Crank and Nicholson \cite{Crank1947} 
for both Hermitian and non-Hermitian type Hamiltonians
\begin{equation} \label{Crank}
\left|v_{n\textbf{k}}\left(t+\triangle t\right)\right\rangle =\frac{I-i\left(\triangle t/2\right)\mathcal{H}_{\textbf{k}}^{system}\left(t\right)}{I+i\left(\triangle t/2\right)\mathcal{H}_{\textbf{k}}^{system}\left(t\right)}\left|v_{nk}\left(t\right)\right\rangle 
\end{equation}
in which $I$ is the identity element. The operation is strictly unitary for any value of time-step $\triangle t$.\\
It turns out that if the applied field is a Dirac delta-type, the Fourier transformed responses can be evaluated at all frequencies. In case of a low intensity, one can show that the solution of Eqn. (\ref{NL_1}) assymtotically tends to Eqn. (\ref{BSE_5}) \cite{Claudio2011}. These are numerically shown in Fig. \ref{fig:fgr6} (a). The extraction of the nonlinear response function is a post-processing computation. We follow Attacallite $et$. $al$. \cite{Claudio2013} for this methodology. As shown in Fig. \ref{fig:fgr6} (c), the sudden switching of a monochromatic $\mathcal{E}(t)$ induces spurious fluctuations at the initial stage. In order to calculate the $\mathcal{P}(t)$ from a clean signal, we add a dephasing time-constant of 7 fs. This would essentially mean that after 5 time-constants ($\sim$32 fs) these spurious fluctuations are sufficiently cleared out from $\mathcal{P}(t)$ and the nonlinear responses can be obtained between this time and the total simulation time ($\sim$55 fs). There are two approaches by which nonlinear $\chi$ can be evaluated. Either, the field may be applied in a quasi-static way, so that the spurious fluctuations do not appear.  However calculating $\mathcal{P}$ in this way takes a long time to simulate \cite{Claudio2013}. 
The other way is to use the previous sudden approximation and change the following Fourier series into a system of linear equations. The polarization Fourier series is \cite{Claudio2013}
\begin{equation} \label{Pol_2}
\mathcal{P}\left(t\right)=\sum_{n=-\infty}^{n=\infty}p_{n}e^{-i\omega_{n}t}
\end{equation}
This series is truncated \cite{Claudio2013} to an order $\mathcal{S}$ larger than the response we are interested to calculate. With a laser frequency $\omega_{L}$, we find the time-period $T_{L}$ and within this we sample the signal to 2$\mathcal{S}+1$ values. Eqn. (\ref{Pol_2}) can now be transformed in a system of linear equation
\begin{equation}
\mathcal{F}_{lin}p_{n}^{\alpha}=\mathcal{P}_{i}^{\alpha}
\end{equation}
in which $\alpha$ is the polarization direction. By Fourier inversion of the $\left(2\mathcal{S}+1\right)\times\left(2\mathcal{S}+1\right)$ matrix (done on sampled times $t_{i}$) $\left(\mathcal{F}_{lin}\equiv e^{-i\omega_{n}t_{i}}\right)$, each component $p_{n}^{\alpha}$ of the coefficients $p_{n}$ can be obtained.

\nocite{*}

\bibliography{apssamp}
\end{document}



\title{Supplemental Material: Exciton-driven giant nonlinear overtone signals from buckled hexagonal monolayer GaAs}
\author{Himani Mishra}
 \author{Sitangshu Bhattacharya}
 \email{Corresponding Author's Email: sitangshu@iiita.ac.in}
\affiliation{Nanoscale Electro-Thermal Laboratory, Department of Electronics and Communication Engineering, Indian Institute of Information Technology-Allahabad, Uttar Pradesh 211015, India}

\newpage 

\maketitle
\onecolumngrid

\section{Supporting convergence figures and added graphical information}
\subsection{Crystal-field splitting and spin-orbit splitting in bulk zincblende and monolayer structure of GaAs.}
\begin{figure}[ht]
  \includegraphics[width=0.8\columnwidth]{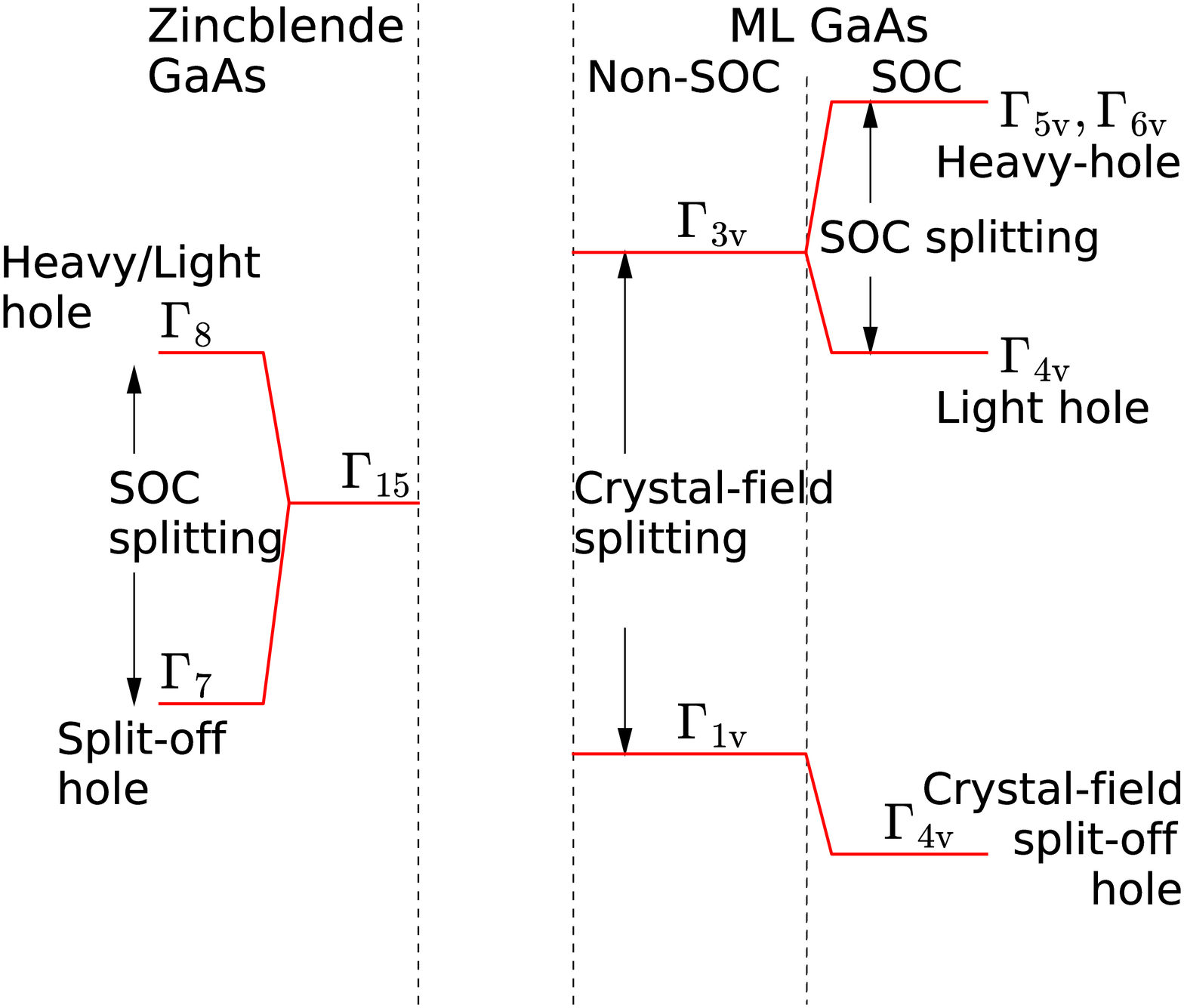}
  \caption{Schematic of valence band splitting at $\Gamma$ of the BZ in the presence of crystal-field and SOC for zincblende and hexagonal GaAs monolayer type structures. Note that $\Gamma_{5\mathrm{v}}$, $\Gamma_{6\mathrm{v}}$ is the Kramers pair with $\left|\mathrm{j},\mathrm{m}_{\mathrm{j}}\right\rangle =\left|\frac{3}{2},\pm\frac{3}{2}\right\rangle $.}
  \label{fig:fgr2}
\end{figure}

\newpage
\subsection{Kinetic cut-off convergences}
\begin{figure*}[ht]
  \includegraphics[width=0.95\columnwidth]{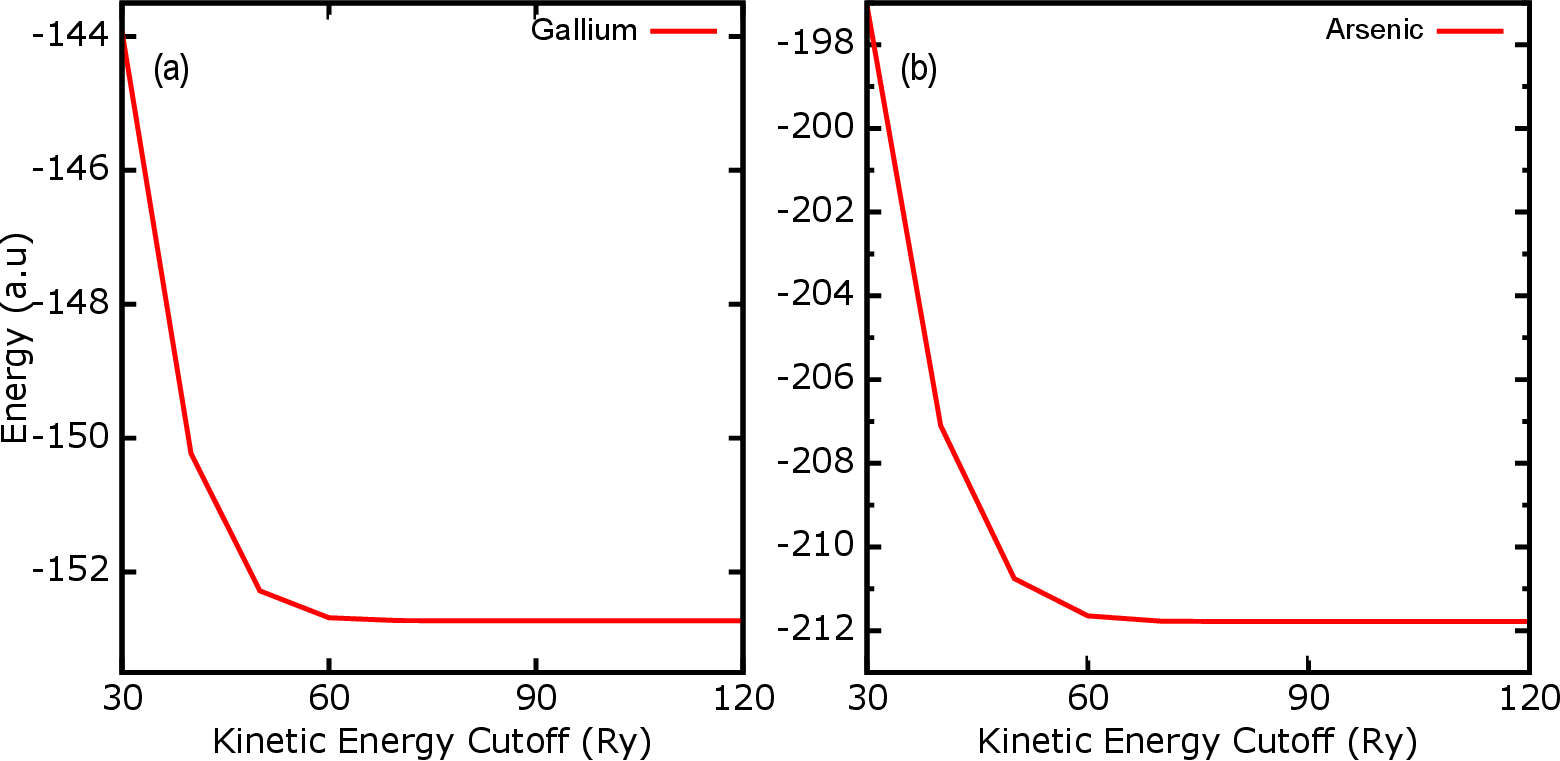}
  \caption{Total energy minimization as function of kinetic energy cut-off using a norm-conserving fully relativistic pseudo-potential of (a) Gallium and (b) Arsenic. We see that the energy settles at 120 Ry in both cases. We therefore choose 120 Ry in all our subsequent computations.}
  \label{fgr:example}
\end{figure*}

\newpage\subsection{k-point sampling convergences in the absence of spin-orbit coupling case}
\begin{figure*}[ht]
  \includegraphics[width=0.7\columnwidth]{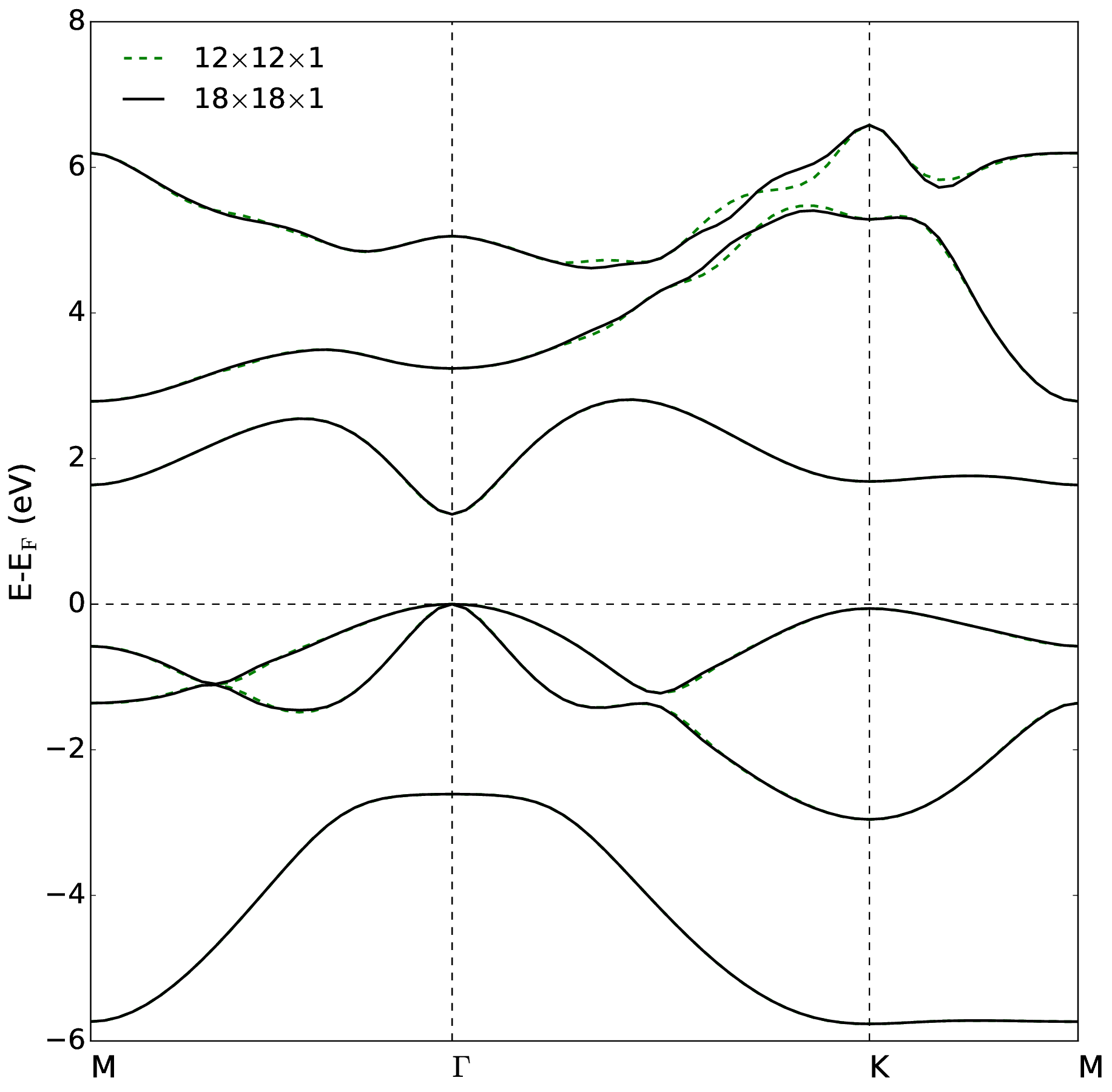}
  \caption{Ground state electron energy dispersion in ML buckled GaAs using kinetic energy cut-of 120 Ry and using a norm-conserving fully relativistic pseudo-potential in absence of spin-orbit coupling for $k$-point sampling of 12$\times$12$\times$1 and 18$\times$18$\times$1. We see that the sampling to 18$\times$18$\times$1 does not improve energies even within meV region, and therefore in all our subsequent DFT analyses, we use 12$\times$12$\times$1 sampling. E$_\mathrm{F}$ is the Fermi energy and the top of the valence band is set to zero energy scale.}
  \label{fgr:example}
\end{figure*}

\newpage\subsection{k-point sampling convergences in the presence of spin-orbit coupling case}
\begin{figure*}[ht]
  \includegraphics[width=0.7\columnwidth]{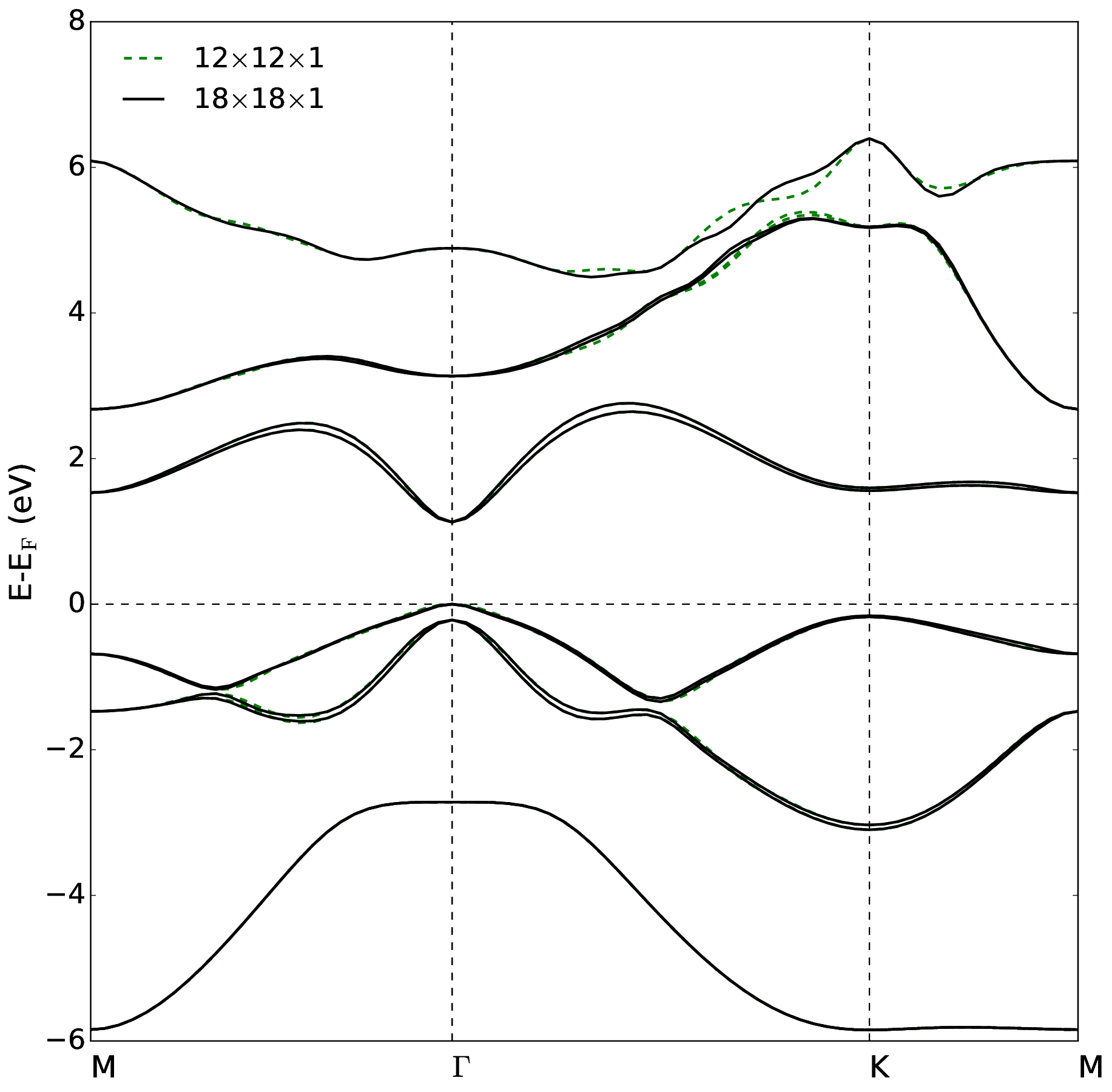}
  \caption{Ground state electron energy dispersion in ML buckled GaAs using kinetic energy cut-of 120 Ry and using a norm-conserving fully relativistic pseudo-potential in presence of spin-orbit coupling for $k$-point sampling of 12$\times$12$\times$1 and 18$\times$18$\times$1. We see that the influence of spin-orbit coupling demonstrate a ``Zeeman"-like splitting in the top of the valence band at $\Gamma$. Dense sampling to 18$\times$18$\times$1 does not improve energies in the conduction and valence bands vicinity to 0-line. Therefore in all our subsequent DFT analyses, we have used 12$\times$12$\times$1 sampling. E$_\mathrm{F}$ is the Fermi energy and the top of the valence band is set to zero energy scale.}
  \label{fgr:example}
\end{figure*}

\newpage\subsection{Energy band structure comparison in the presence and absence of spin-orbit coupling}
\begin{figure*}[ht]
  \includegraphics[width=1\columnwidth]{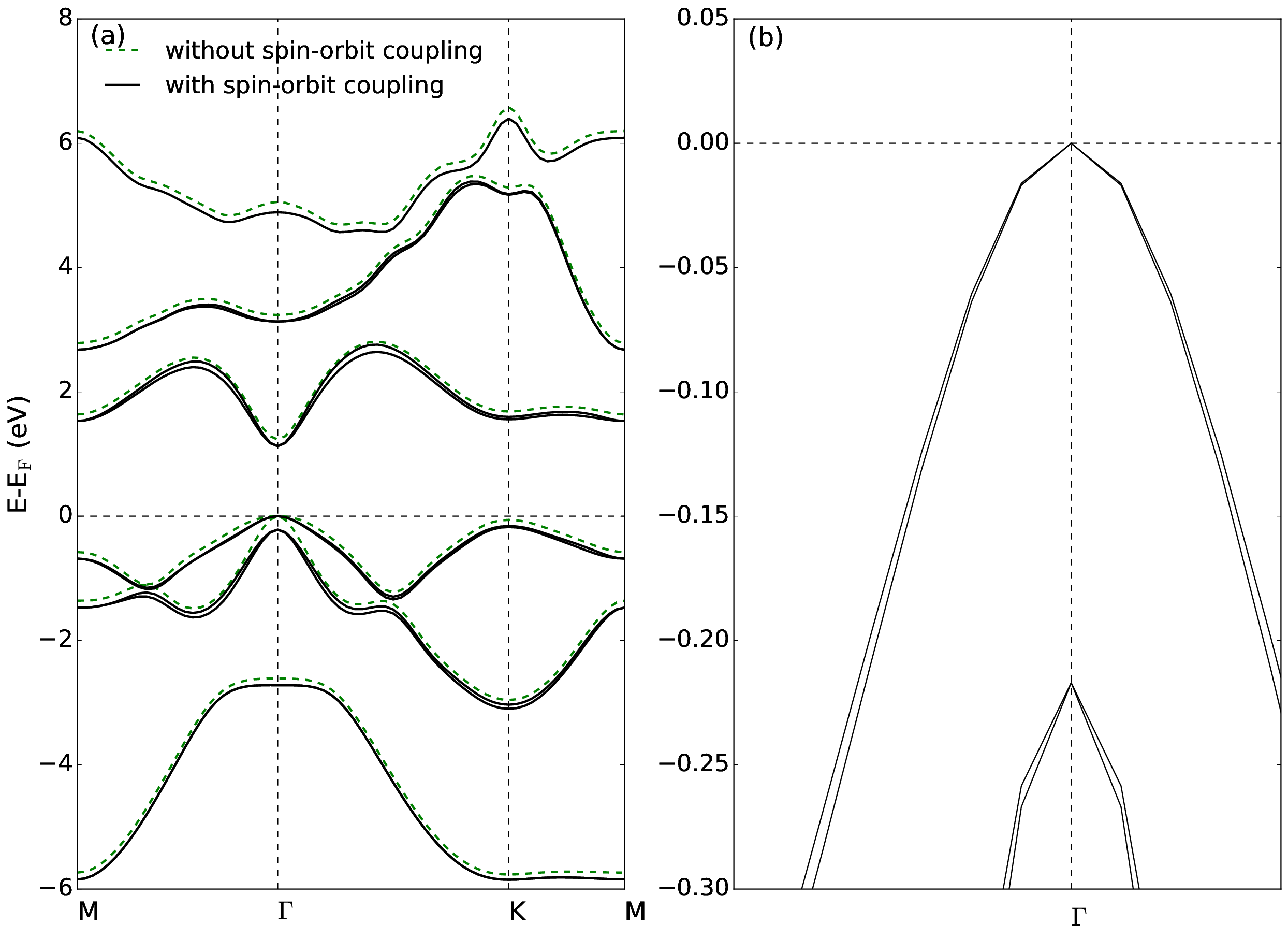}
  \caption{(a) Ground state electron energy dispersion in ML buckled GaAs in the presence and absence of spin-orbit coupling. We see that the influence of spin-orbit coupling demonstrate a ``Zeeman"-like splitting in the top of the valence band at $\Gamma$. E$_\mathrm{F}$ is the Fermi energy and the top of the valence band is set to zero energy scale. (b) Magnified snap-shot of the degeneracy in the valence states at $\Gamma$ in the presence of spin-orbit coupling.}
  \label{fgr:example}
\end{figure*}

\newpage\subsection{Phonon dispersion convergences}
\begin{figure*}[ht]
  \includegraphics[width=0.8\columnwidth]{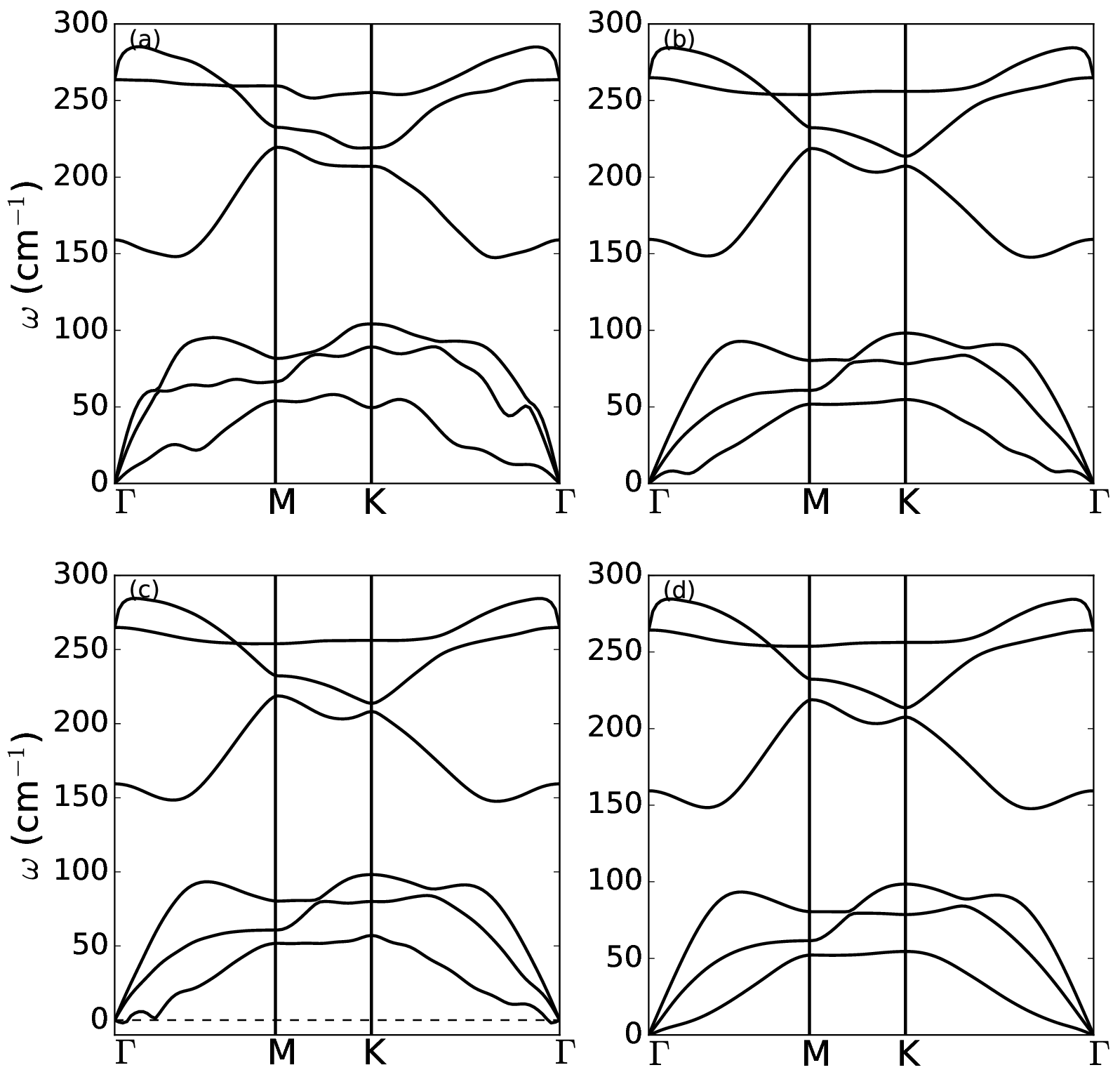}
  \caption{Phonon dispersion curves showing the convergences using (a) 12$\times$12$\times$1 phonon grid and a rigid self-consistent
error threshold below 10$^{-12}$ Ry, (b) 14$\times$14$\times$1 phonon grid and a rigid self-consistent
error threshold below 10$^{-14}$ Ry, (c) 18$\times$18$\times$1 phonon grid and a rigid self-consistent
error threshold below 10$^{-16}$ (soft-modes are seen in  the vicinity of $\Gamma$ along $\Gamma$-\textbf{M} route) Ry and (d) 18$\times$18$\times$1 phonon grid and a rigid self-consistent
error threshold below 10$^{-18}$ Ry. Clearly, (d) shows a converged result with no soft-modes for the geometry in Fig. S1.}
  \label{fgr:example}
\end{figure*}

\newpage\subsection{BSE spectra convergence with respect to local field effect cut-off}
\begin{figure*}[ht]
  \includegraphics[width=0.95\columnwidth]{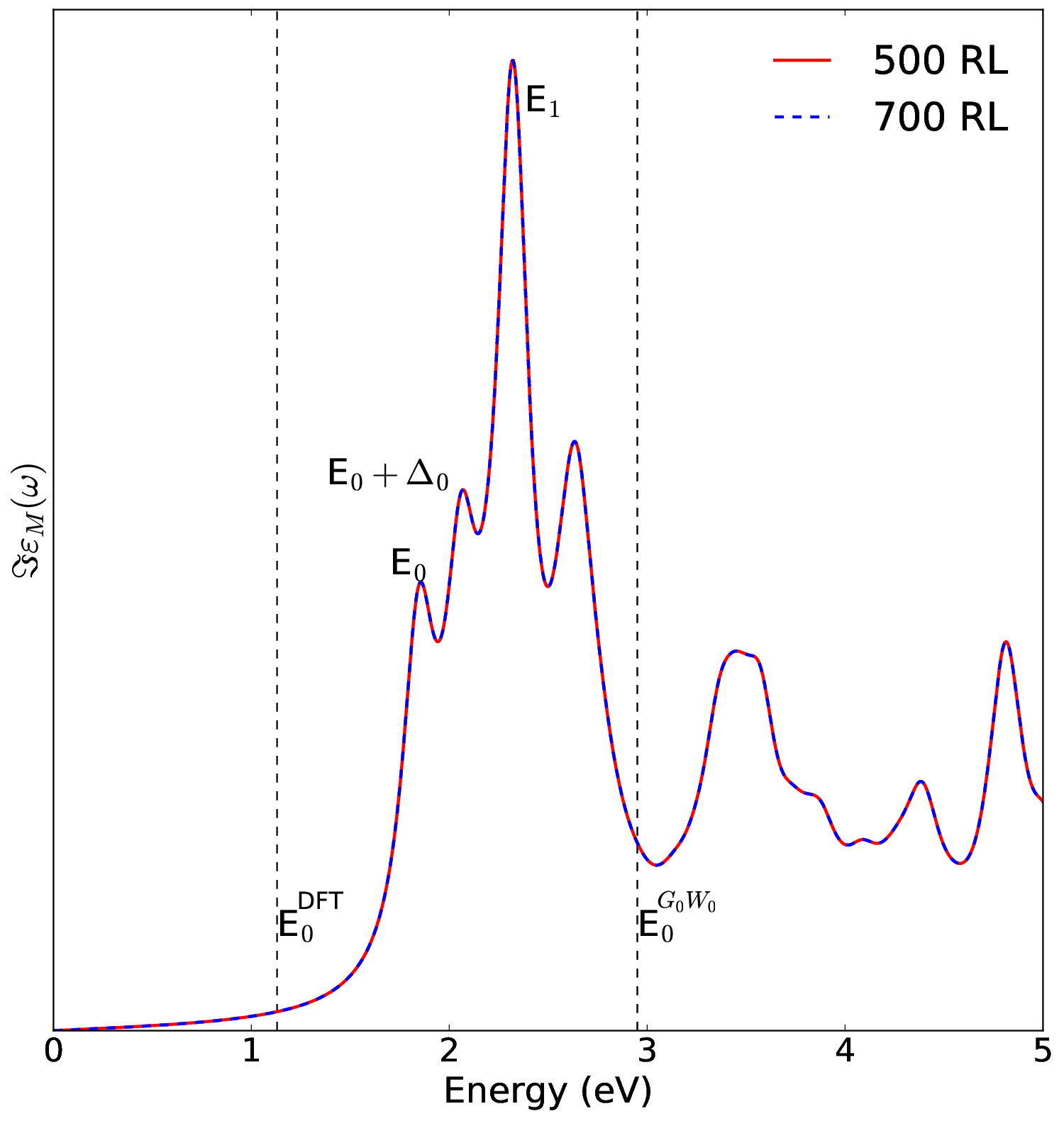}
  \caption{Comparison between the absorption spectra for two different values of the response block size used in calculation of local field effects (LFEs). As optical spectra are very sensitive to the $k$-point sampling, we perform all our static and time-dependent spectra on 72$\times$72$\times$1 $k$-point grid. Here the cut-off energies are in the unit of reciprocal lattice (RL) vectors. We see that the convergence is achieved for both 500 RL ($\sim$7 Ry) and 700 RL ($\sim$9 Ry).}
  \label{fgr:example}
\end{figure*}

\newpage\subsection{BSE absorption spectra convergence with respect to transition bands}
\begin{figure*}[ht]
  \includegraphics[width=0.8\columnwidth]{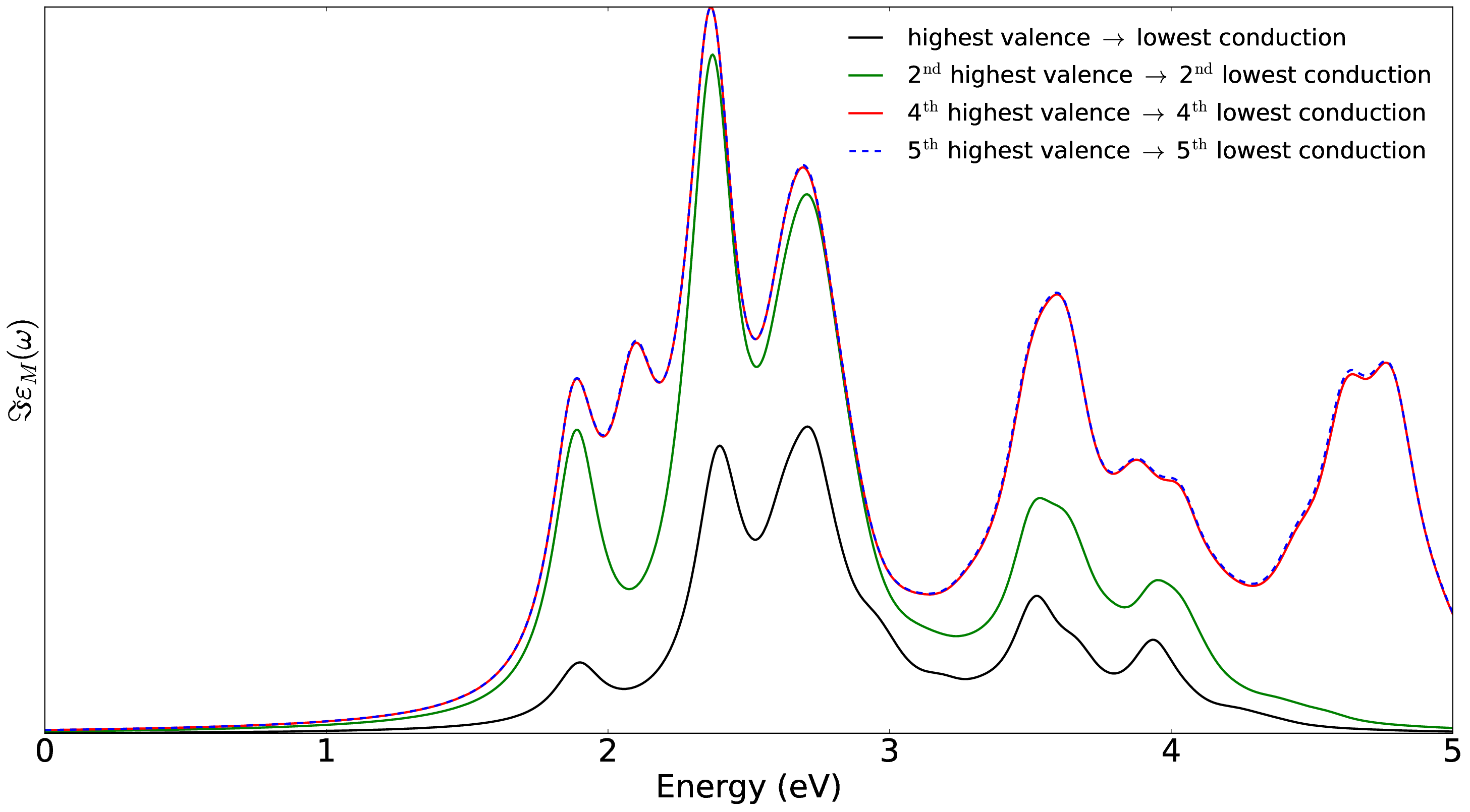}
  \caption{Convergence of absorption spectra with respect to the number of conduction and valence bands. The local field effect cut-off of 7 Ry is used from the preceding Fig. S7. As optical spectra are very sensitive to the $k$-point sampling, we perform all our static and time-dependent spectra on 72$\times$72$\times$1 $k$-point grid. An excellent convergence is shown using top 5 valence and  lowest 5 conduction bands.}
  \label{fgr:example}
\end{figure*}

\newpage\subsection{SHG absorption spectra of monolayer and bulk zincblende structures of GaAs.}
\begin{figure*}[ht]
  \includegraphics[width=1.0\columnwidth]{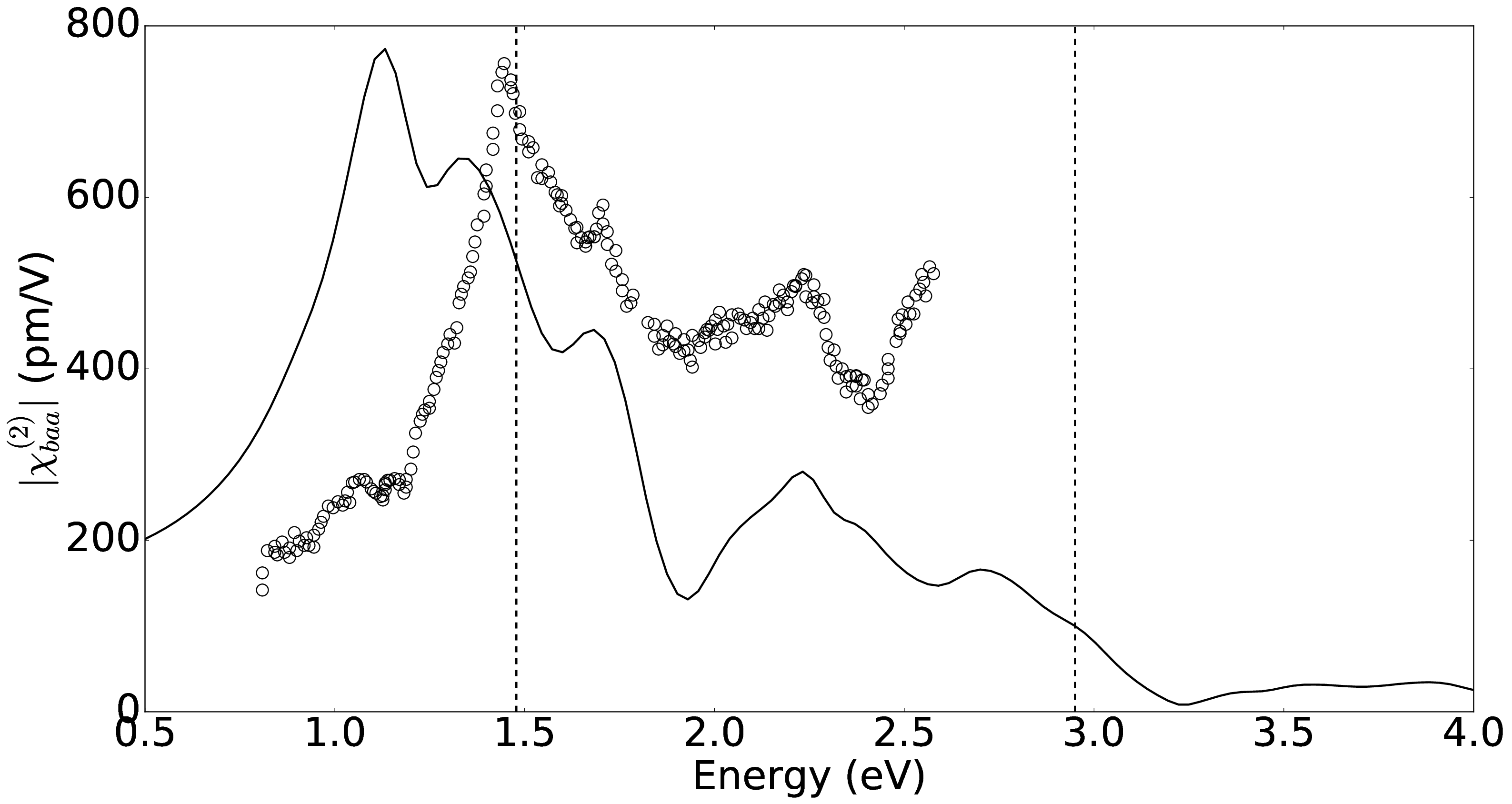}
  \caption{Nonlinear SHG spectra of ML buckled GaAs and bulk zincblende GaAs. The solid line is the SHG computed using the TD-BSE level of theory. The symbols are the experimental SHG spectra of zincblende bulk GaAs taken from Bergfeld and Daum \cite{Bergfeld2003} and put here for comparison with the hexagonal monolayer. The solid and dashed vertical lines are for the $\omega$ and $\omega/2$ gaps of ML GaAs. The theoretical computations were performed on 72$\times$72$\times$1 $k$-point grid.}
  \label{fgr:example}
\end{figure*}

\newpage\subsection{Individual atomic partial density-of-states on the energy band of ML GaAs.}
\begin{figure*}[ht]
  \includegraphics[width=1.0\columnwidth]{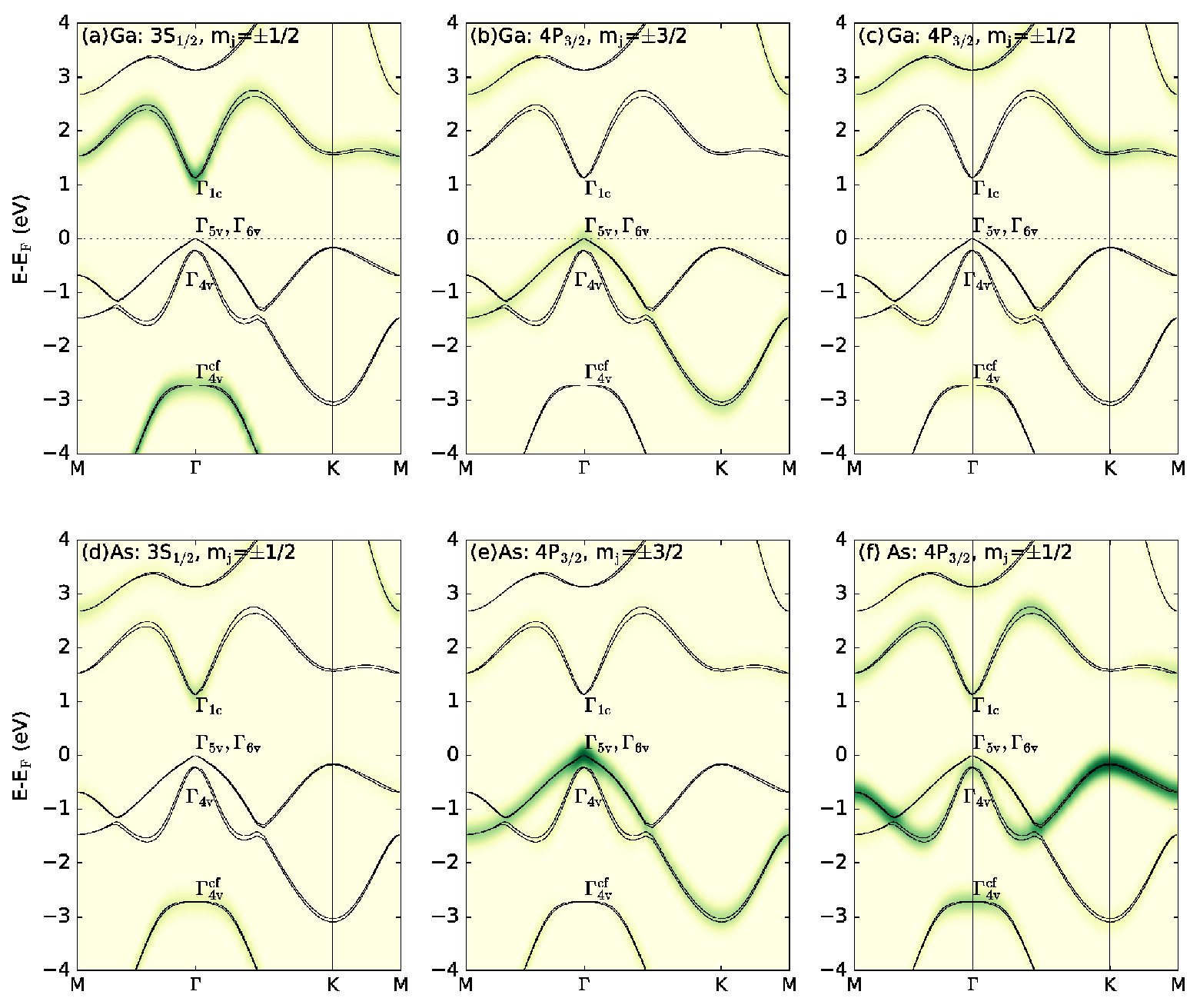}
  \caption{Ground state electron energy dispersion in ML buckled GaAs in the presence SOC demonstrating the individual p-DOS contributions from (a) Ga 3S$_{1/2}$, m$_\mathrm{j}$=1/2, (b) Ga 4P$_{3/2}$, m$_\mathrm{j}$=$\pm$3/2, (c) Ga 4P$_{3/2}$, m$_\mathrm{j}$=$\pm$1/2, (d) As 3S$_{1/2}$, m$_\mathrm{j}$=1/2, (e) As 4P$_{3/2}$, m$_\mathrm{j}$=$\pm$3/2 and (f) As 4P$_{3/2}$, m$_\mathrm{j}$=$\pm$1/2. We find that at $\Gamma$, the orbital weight of the state $\Gamma_{1\mathrm{c}}$ and $\Gamma_{4\mathrm{v}}^{\mathrm{cf}}$ comes from Ga 3S$_{1/2}$, m$_\mathrm{j}$=1/2. The valence spin-orbit splitting ($\Delta_{0}=\Gamma_{5\mathrm{v}} -\Gamma_{4\mathrm{v}}$) and the direct band-gap (E$_0=\Gamma_{1\mathrm{c}} -\Gamma_{5\mathrm{v}}$) at $\Gamma$ is 0.22 eV and 1.13 eV respectively. E$_\mathrm{F}$ is the Fermi energy and the top of the valence band is set to zero energy scale.}
  \label{fgr:example}
\end{figure*}

%
\newpage\bibliography{apssamp_supple}